\documentclass[5p,12pt]{elsarticle}
\biboptions{sort&compress}

\usepackage[utf8]{inputenc}
\usepackage[T1]{fontenc}
\usepackage[fleqn]{amsmath}
\usepackage{amssymb}
\usepackage{amsfonts}
\usepackage{xcolor}
\usepackage{hyperref}
\usepackage{graphicx}
\usepackage{amsthm}
\usepackage{array}
\usepackage{bm}
\usepackage{colortbl}
\usepackage{tabularx}

\usepackage{subcaption}
\usepackage{comment}

\usepackage{pgfplots}
\pgfplotsset{compat=1.9}
\usepackage[font=small,labelfont=bf,justification=centering]{caption}

\usepackage{tikz}
\pgfkeys{/pgf/number format/set thousands separator={}}

\usetikzlibrary{shapes}

\colorlet{mydarkblue}{blue!40!black}
\colorlet{mylightblue}{mydarkblue!12} 
\colorlet{myred}{red!80!black}
\colorlet{mydarkred}{red!50!black}
\colorlet{mylightred}{mydarkred!12}
\colorlet{mydarkgreen}{red!60!black!70}
\colorlet{mylightgreen}{mydarkgreen!15}
\colorlet{myorange}{orange!63!black}
\colorlet{mylightorange}{orange!80!black!12}

\colorlet{gradienttopleft}{orange!70!black!25}
\colorlet{gradientbottomright}{violet!70!black!25}

\usetikzlibrary{calc, decorations, patterns, arrows, decorations.pathmorphing}
\definecolor{pltblue}{HTML}{1F77B4}
\makeatletter
\pgfset{
	/pgf/decoration/randomness/.initial=2,
	/pgf/decoration/wavelength/.initial=100	
}
\pgfdeclaredecoration{sketch}{init}{
	\state{init}[width=0pt, next state=draw, persistent precomputation={
		\pgfmathsetmacro\pgf@lib@dec@sketch@t0	
	}]{}

	\state{draw}[width=\pgfdecorationsegmentlength,
	auto corner on length=\pgfdecorationsegmentlength,
	persistent precomputation={
		\pgfmathsetmacro\pgf@lib@dec@sketch@t{
			mod(\pgf@lib@dec@sketch@t + pow(2,rand), 100)
		}	
	}
	]
    {
		\pgfmathparse{sin(2*\pgf@lib@dec@sketch@t*pi/100 r)}
		\pgfpathlineto{\pgfqpoint{\pgfdecorationsegmentlength}{\pgfmathresult\pgfdecorationsegmentamplitude}}
	}

	\state{final}{}
}
\tikzset{xkcd/.style={decorate, decoration={sketch, segment length=0.5pt, amplitude=0.5pt}}}
\makeatother
\pgfmathsetseed{1}
\usetikzlibrary{shapes.geometric, arrows}

\tikzstyle{startstop} = [rectangle, rounded corners, 
minimum width=3.1cm, 
minimum height=1cm,
text centered, 
draw=black, 
fill=red!30]

\tikzstyle{io} = [trapezium, 
trapezium stretches=true,
trapezium left angle=70, 
trapezium right angle=110, 
minimum width=3.1cm, 
minimum height=1cm, text centered, 
draw=black, fill=blue!30]

\tikzstyle{iohigh} = [trapezium, 
trapezium stretches=true,
trapezium left angle=70, 
trapezium right angle=110, 
minimum width=3.1cm, 
minimum height=1cm, text centered, 
draw=black,
fill=red!50]

\tikzstyle{process} = [rectangle, 
minimum width=3.1cm, 
minimum height=1cm, 
text centered, 
text width=3cm, 
draw=black, 
fill=orange!30]

\tikzstyle{highcost} = [rectangle, 
minimum width=3.1cm, 
minimum height=1cm, 
text centered, 
text width=3cm, 
draw=black, 
fill=red!50]

\tikzstyle{nocost} = [rectangle, 
minimum width=3.1cm, 
minimum height=1cm, 
text centered, 
text width=3cm, 
draw=black]

\tikzstyle{decision} = [diamond, 
minimum width=2.7cm, 
minimum height=1cm, 
text centered, 
draw=black, 
fill=green!30]

\tikzstyle{arrow} = [thick,->,>=stealth]
\AtBeginEnvironment{tikzpicture}{\tracinglostchars=0\relax}

\newcounter{bla}

\journal{Computer Physics Communications}

\begin{document}

\begin{frontmatter}

\title{\textsc{Archê}, an orbital-free molecular dynamics code \\for fast production of equations of state}

\author[a,b]{William~Weens\corref{author}}
\author[a,b]{Augustin~Blanchet}
\author[a,b]{Philippe~Arnault}

\cortext[author] {Corresponding author.\\\textit{E-mail address:} william.weens@cea.fr}
\address[a]{CEA, DAM, DIF, F-91297 Arpajon, France}
\address[b]{Université Paris-Saclay, CEA, Laboratoire Matière en Conditions Extrêmes, 91680 Bruyères-le-Châtel, France}

\begin{abstract}
We present \textsc{Archê}, an orbital-free molecular dynamics (OFMD) code designed to produce equations of state (EOS) in the plasma state. Unlike in other OFMD codes, \textsc{Archê} uses a self-consistent field (SCF) approach to compute the electronic density. This allows us to implement two algorithms that accelerate SCF convergence by a factor of up to six. First, the density is initialized using the results from the previous MD timestep to define a one-center profile, which is then applied to the new nuclei positions. Second, the initial and final densities are mixed at each SCF iteration in a proportion that minimizes an approximate free energy. We validate the code by comparing the calculated aluminum EOS to results obtained with the Kohn-Sham density functional theory software \textsc{Abinit}. Achieving agreement in the internal energy requires adding a correction related to the norm-conserving pseudopotential derived from an average-atom model. Performance is compared across CPU and GPU architectures, demonstrating an order-of-magnitude speedup for a single GPU compared to 256 CPUs. \textsc{Archê} exhibits an overall linear computational complexity with respect to the number of atoms, as well as the number of real and reciprocal grid points. Execution time is weakly dependent on density; however, interestingly, it decreases as temperature increases—in contrast to simulations based on Kohn-Sham orbitals.
\end{abstract}

\end{frontmatter}

\section{Introduction}
\label{intro}
Many fields have benefited from Kohn-Sham density functional theory (KSDFT) for electronic structure calculations \cite{Burke2012PerspectiveOD}. Thousands of papers using KSDFT are published every year. Despite the progress made in computing science since the method first appeared, \textit{ab initio} calculations for large systems or at high temperatures remain elusive, costly, and sometimes out of reach. This limitation primarily stems from the orthogonalization of the wave functions (also called orbitals\footnote{Orbital is the term used "to denote a solution of the wave equation for a system of only one electron". A system of $N$ electrons can be approximated "by assigning the $N$ electrons to $N$ different orbitals, where each orbital is a solution of a wave equation for one electron" \cite{Kittel2004}.}) associated with the energy levels of electrons \cite{Kresse1996EfficiencyOA}. As a rule of thumb, the calculation scales as $N^3$, where $N$ is the number of orbitals. The more atoms there are to simulate, the more electrons are present, and the more orbitals must be computed. At finite temperatures, the situation is exacerbated because electrons occupy higher energy levels, resulting in significantly more orbitals to compute \cite{Blanchet2020}.

This scaling issue prevents extensive applications of KSDFT. Indeed, many physical situations of interest require simulating a very large number of atoms. This need is driven by the high structural complexity of molecular chemistry and materials physics (see \cite{Yin2021AtomisticSO} and references therein). Phenomena outside local thermodynamic equilibrium conditions, such as mass diffusion at an interface or non-Newtonian fluids, also demand large simulation cells. Conversely, while only a few fields require an accurate account of temperature effects on electronic structure, they are crucial for advancing our knowledge. Astrophysics is one of them, encompassing a vast array of conditions ranging from giant planets to stellar interiors and supernova explosions \cite{Remington2004ExperimentalAW}. Temperature-dependent electronic structure is also critical in another, more technological field that has recently seen impressive breakthroughs: inertial confinement fusion (ICF) \cite{Lindl1995DevelopmentOT,AbuShawareb2024AchievementOT,ICF2022}.

These fields share a common need for hydrorad simulations. Consequently, there is a constant demand for new data to describe material behavior, including equations of state (EOS), opacities, and transport coefficients. In the context of ICF, these quantities must be provided on short timescales to keep pace with rapid technological advances. They are required over an extremely wide range of conditions, spanning densities from $10^{-5}$ to $10^5$ g/cm$^3$ and temperatures from $300$ K to $10^9$ K, including the particularly challenging regime known as warm dense matter (WDM) \cite{Bonitz2019AbIS}.

Accurate data at these very high temperatures can be obtained thanks to recent and ongoing efforts to extend KSDFT applications to higher temperatures \cite{Suryanarayana2017SQDFTSQ, ExtendedZhang2016, BlanchetAluminum2022, Baer2013SelfaveragingSK, White2020FastAU, Liu2022PlanewavebasedSD, Sharma2023StochasticAM}. Unfortunately, these advances require significant computing resources, \linebreak which are incompatible with the rapid turnaround required in the ICF technology cycle between experiments and simulations. In the present work, we use one of these KSDFT extensions as a reference. This extension bypasses the $N^3$ scaling bottleneck by replacing the highest orbitals with plane waves excluded from the orthogonalization computation \cite{ExtendedZhang2016}. This approach is implemented as an orbital-based option in the \textsc{Abinit} code called \textsc{Extended DFT} (originally known as \textsc{Extended FPMD}) \cite{Blanchet2022PhD,BlanchetAluminum2022}. Other alternatives have also been proposed and are currently being tested \cite{Suryanarayana2017SQDFTSQ, Baer2013SelfaveragingSK, White2020FastAU, Liu2022PlanewavebasedSD, Sharma2023StochasticAM}.

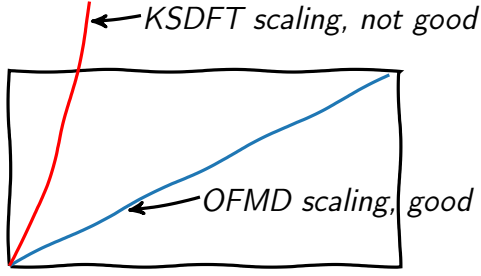
\begin{figure}
\makebox[0.8\columnwidth][c]{
\begin{tikzpicture}[xscale=1.2, yscale=1.2]
	\node[text width=10.3cm, align=right] (S) at (1.9, 3.2) {\fontfamily{cmss}\selectfont\textit{KSDFT scaling, not good}};
	\node[text width=8.0cm, align=right] (N) at (2.8, 1.2) {\fontfamily{cmss}\selectfont\textit{OFMD scaling, good}};
	\begin{scope}[very thick, yscale=0.5, every path/.style={xkcd}]	
	\draw (1.,1.) rectangle (5.3,5.3);
	\draw[domain=1.0:5.2, pltblue, samples=100] plot (\x, {\x});
	\draw[domain=1.0:1.9, red, samples=100] plot (\x, {\x^3});
	\draw[->, >=stealth'] ($(S.north)-(-0.55cm, 0.45cm)$) -- (1.89, 6.4);
	\draw[->, >=stealth'] ($(N.north)-(-0.3cm, 0.45cm)$) -- (2.3, 2.2);
	\end{scope}
	
\end{tikzpicture}

}
\caption{Scaling behavior with respect to system size, sketched as in \cite{Mi2023}. Due to the orthogonalization of orbitals, KSDFT scaling is $\sim \mathcal{O}(N^3)$, whereas OFDFT scaling is linear $\sim \mathcal{O}(N)$.}
\label{fig:complexity0}
\end{figure}

To meet the material property demands of ICF, a fair trade-off must be struck between accuracy and efficiency. Another way to circumvent the KSDFT scaling issue is to use orbital-free density functional theory (OFDFT) coupled with Born-Oppenheimer molecular dynamics (OFMD) \cite{Mi2023, Karasiev_2025}. The OFDFT approach relies primarily on a local density approximation for the kinetic energy \cite{Feynman1949}. With this approximation, orbitals are no longer needed. This speeds up calculations while maintaining accuracy, at least at high temperatures. For low temperatures, a more refined approximation of the non-interacting kinetic energy functional is required to restore accuracy. Semi-local corrections involving the density gradient exist \cite{Luo2020}, as well as non-local functionals utilizing a kernel that connects the density at two different points \cite{RiosVargas2024, Bhattacharjee2024}. These investigations into kinetic energy functionals parallel those for exchange and correlation functionals \cite{Perdew2001JacobsLO}. With OFDFT, since only the electron density is computed instead of many orbitals, the computation time scales as $N$ rather than $N^3$ (see Fig.\,\ref{fig:complexity0}). As noted in \cite{MIHAYLOV2024108931}, the advantages of a fast code expand the domain of DFT applications (e.g., reaching higher temperatures and improving statistics). It opens up new possibilities, such as preprocessings of KSDFT simulations to reach equilibrium configurations, rapid generation of low-fidelity data for machine learning \cite{Tran2020}, and the calculation of dynamic properties (like viscosity) that require many atoms and timesteps \cite{Clerouin2020}.

Codes implementing OFDFT can be categorized by their treatment of temperature effects. For zero-$T$ conditions and large systems, \textsc{profess} \cite{Profess2008} and \textsc{DFTpy} \cite{Shao2020DFTpyAE} implement periodic boundary conditions, but real-space implementations also exist to handle more general boundary conditions \cite{Mi2015ATLASAR, Atlas2024, Lehtomki2014OrbitalFreeDF, Golub2020CONUNDrumAP, Gavini2007QuasicontinuumOD, Suryanarayana2014AugmentedLF, Das2015RealspaceFO}. Unfortunately, these codes cannot be utilized for ICF applications, where temperature effects are paramount. At finite temperatures, there are fewer OFMD codes available: \textsc{profess}@\textsc{qe} \cite{KARASIEV20143240}, \textsc{ofmd} \cite{Lambert2006}, and more recently \textsc{dragon} \cite{MIHAYLOV2024108931}.

In this paper, we present \textsc{Archê}, a finite-temperature OFMD code. The core design of the code is based on the approach described in \cite{Lambert2007}, with several notable differences: the use of more recent programming languages and paradigms;\linebreak GPU support; the inclusion of unit and integration tests; a scalable three-dimensional Fourier transform; the adoption of a self-consistent field (SCF) method instead of free energy minimization via a conjugate-gradient algorithm (which, to our knowledge, is employed by all existing OFDFT codes); initialization of the electronic density using the converged density from the previous MD timestep and a density-mixing strategy following \cite{More1988}, which both significantly accelerate SCF convergence. Some of these developments arise from general physical considerations and can be readily transferred to other modeling frameworks.

\textsc{Archê} is a parallel code written in C++,\linebreak CUDA, and Python. It is linked to \textsc{HeFFTe} \cite{HEFFTEfirst, HEFFTElast}, a scalable fast Fourier transform (FFT) library. This code serves a dual purpose: first, to quickly produce EOS for WDM; second, to explore numerical methods and implementations that enhance performance, as well as to test other non-interacting free energy functionals to improve accuracy \cite{Luo2020, WangTeter1992}. A particular effort has been made to build a flexible and modular framework so that ideas can be tested without compromising performance. Following a free-function–based programming approach \cite{Stroustrup2014}, all energy contributions and functionals are implemented as independent functions that can be freely combined or replaced. To substantiate the presentation of the code, we detail its validation process and assess its performance on both CPU and GPU architectures.

The article is organized as follows. Section \ref{model} summarizes the OFMD theoretical framework. Section \ref{implementation} describes its implementation in \textsc{Archê}. Section \ref{validation} validates the code by comparing its results with more accurate models. Finally, Section \ref{benchmark} highlights the performance of \textsc{Archê} through various benchmarks.
\section{Theoretical background}
\label{model}

In this section, we present the model as implemented in the \textsc{Archê} code. Unless otherwise stated, equations are written in atomic units ($e = 4 \pi \varepsilon_0 = \hbar = m_e = 1$). 

\subsection{Molecular dynamics, the main loop}
The purpose of a molecular dynamics simulation is to evolve a system of $N_i$ atoms over time. For each atom, this is achieved by solving Newton's equation for the force $\vec{F}_j$ acting on the atomic nucleus $j$:
\begin{equation}
\vec{F}_j = m_j \vec{a}_j,
\end{equation}

\noindent where $m_j$ is the mass of the nucleus and $\vec{a}_j$ is its acceleration.

The specificity of each MD model lies in the accuracy or scale at which the forces are computed. The common aspect in MD models is time discretization. There are several schemes, each with its own properties and an associated statistical ensemble in which the simulation is performed \cite{Martin2020, MecaStatsTuckerman}. We chose the Verlet integration algorithm, which enables simulations in the microcanonical ensemble (the isokinetic ensemble algorithm, also implemented in \textsc{Archê}, follows similar lines \cite{Tuckerman2003b}).

The Verlet scheme consists of two steps. The first step determines the new position of the nucleus $\vec{r}_j$ at time $t + \Delta t$, where $\Delta t$ is the MD timestep of the simulation:

\begin{equation}
\vec{r}_j(t + \Delta t) = 2 \vec{r}_j(t) - \vec{r}_j(t - \Delta t) + \frac{\vec{F}_j(t)}{m_j} \Delta t^2\, .
\end{equation}

The second step computes the velocities $\vec{v}_j$ at time $t$ from the positions at times $t + \Delta t$ and $t - \Delta t$:
\begin{equation}
\vec{v}_j(t) = \frac{\vec{r}_j(t + \Delta t) + \vec{r}_j(t - \Delta t)}{2 \Delta t}\, .
\end{equation}

These steps are sufficient to evolve the system, provided that the forces are known. Since the propagating particles are charged nuclei, the forces $\vec{F}_j$ that set the $N_i$ nuclei in motion are of Coulombic origin. These forces can be broken down into two contributions, $\vec{F}_j^{ii}$ and $\vec{F}_j^{ie}$, which are derived from the corresponding potentials: $U^{ii}$ and $U^{ie}$ for the nucleus-nucleus and nucleus-electron interactions, respectively.

At each time $t$, each $\vec{F}_j(t)$ depends on the positions of the entire system $R(t) = (\vec{r}_1(t), \ldots, \vec{r}_{N_i}(t))$ and is given by
\begin{eqnarray}
\label{eq:force}
&\vec{F}_j(t) &= \vec{F}_j^{ii}(t) + \vec{F}_j^{ie}(t)   \\
& &= -\nabla_{\vec{r}_j}(U^{ii}(R(t)) + U^{ie}(R(t))).\nonumber
\end{eqnarray}
In the following, we assume the Born-Oppenheimer approximation, in which the electronic response is instantaneous, ensuring that the electronic density is in equilibrium with the nuclear positions at any given time \cite{Paquet2018}. The electron-nucleus interaction $U^{ie}$ uses the electronic density, computed within the density functional theory framework summarized below. The nucleus-nucleus interaction $U^{ii}$ is evaluated using Ewald summation (see Sec.\ref{sec:thermo}). 

\subsection{Orbital-free model}
The electronic density $n$ is a continuous field over the domain $\Omega$ that contains a system of $N_e$ electrons, defined as
\begin{equation}
\label{defN}
N_e = \int_\Omega n(\vec{r}) d\vec{r}.
\end{equation}

Hohenberg and Kohn proved in 1964 \cite{ParrYang1989} that for an external potential $V_{\text{ext}}(\vec{r})$, there exists a functional $\mathcal{E}[n]$ that yields the total ground-state energy $E_0$ via:
\begin{equation}
\label{minE}
E_0 = \underset{n}{\text{min}}\left(\mathcal{E}[n] + E_{\text{ext}}[n]\right).
\end{equation}

\noindent where $E_{\text{ext}}[n] = \int_\Omega n(\vec{r}) V_{\text{ext}}(\vec{r}) d\vec{r}$ represents the energy of the electrons in the potential $V_{\text{ext}}$.
 
When solving Eq.(\ref{minE}), KSDFT uses orbitals (which are more precise but computationally expensive), whereas OFDFT models work solely with $n$ (see the comprehensive review in \cite{Mi2023} for a detailed comparison). In both cases, $n$ must integrate to $N_e$, and the problem is equivalent to extremizing the Lagrangian subject to the constraint:
\begin{eqnarray*}
&\mathcal{L}[n(\vec{r})] = &\mathcal{E}[n(\vec{r})] + E_{\text{ext}}[n(\vec{r})] \\
& &- \mu \left(\int n(\vec{r}) d\vec{r} - N_e\right),\nonumber
\end{eqnarray*}
 where $\mu$ represents the chemical potential, \textit{i.e.}, the Fermi energy at $T=0$.

At finite temperatures, the energy functional $\mathcal{E}$ is replaced by a temperature-dependent functional $\mathcal{F}$, and the free energy $F$ is obtained by extremizing the following Lagrangian \cite{Mermin1965}:
 \begin{eqnarray}
  \label{Lagrangian}
& \mathcal{L}[n(\vec{r})] = &\mathcal{F}[n(\vec{r}), T] + E_{\text{ext}}[n(\vec{r})] \\
& &- \mu \left(\int n(\vec{r}) d\vec{r} - N_e\right).\nonumber
 \end{eqnarray}

$\mathcal{F}$ can be broken down into three different contributions: a non-interacting free-energy functional $\mathcal{F}_0$, the Hartree energy functional $E_H$, and the exchange and correlation free-energy functional $\mathcal{F}_{\text{xc}}$ \cite{Perrot1979, Karasiev_2025}. Minimizing Eq.~\eqref{Lagrangian} with respect to $n$ leads to the following Euler equation:
\begin{equation}
  \label{Eulerian}
\frac{\delta \mathcal{F}_0}{\delta n} + V_{\text{eff}} = \mu 
\end{equation}
\noindent with $V_{\text{eff}}:= V_H + V_{\text{xc}} + V_{\text{ext}}$, where $V_{\text{xc}}:= \frac{\delta \mathcal{F}_{\text{xc}}}{\delta n}$ and $V_H := \frac{\delta E_H}{\delta n} = \int_\Omega \frac{n(\vec{r}\,')}{|\vec{r} - \vec{r}\,'|} d\vec{r}\,'.$

A century ago, Thomas and Fermi independently found an analytic solution to Eq.(\ref{Eulerian}) for an ideal gas of non-interacting electrons in a mean-field \textit{effective} potential $V_{\text{eff}}$ (see Ref. \cite{Nikiforov2005} for the derivation):
\begin{equation}
 \label{SCF1}
 n(\vec{r}, \mu) = \frac{(2 k_B T_e)^{3/2}}{2 \pi^2} I_{1/2} \left( \frac{\mu - V_{\text{eff}} (\vec{r}) }{k_B T_e}\right) .
\end{equation}
Here, $I_{1/2}(.)$ is the Fermi-Dirac integral of order one-half defined as $ \operatorname {I} _{j}(x) = \int _{0}^{\infty }\!{\frac {t^{j}}{\mathrm {e} ^{t-x}+1}}\;\mathrm {d} t$, $k_B$ is the Boltzmann constant, and $T_e$ is the electronic temperature. For EOS purposes, $V_{\text{ext}} := V^{\text{ie}}$, where $V^{\text{ie}}$ is the Coulomb potential of the nuclei, and $V_{\text{H}}$, denoted as $V^{\text{ee}}$, is the Hartree potential solution to the Poisson equation:
\begin{equation}
 \label{SCF2}
\Delta V^{\text{ee}}(\vec{r}) = -4 \pi n(\vec{r}, \mu).
\end{equation}

Within the Thomas-Fermi (TF) model, two main sources of error can be identified. The first arises from the assumption of a uniform electron gas for the functional $\mathcal{F}_0$, while the second is associated with the mean-field approximation. Several refinements of this model exist, including the von Weizsäcker gradient correction to the uniform electron gas \cite{vonWeizsacker1935} and exchange-correlation effects with\-in the mean-field approximation \cite{xcKSDT}.

\subsection{Discretization of the periodic cell}
\label{convention:fourier}

Exploiting the three-dimensional periodicity of the system, the Poisson equation is solved in Fourier space using FFTs, which requires a regular grid representation in both real and reciprocal spaces.

We study a periodic system composed of an infinite repetition of the unit cell $\Omega$. In the real spatial domain, it is a box of dimensions $[L_x \times L_y \times L_z]$, and $\widehat{\Omega}$ is the reciprocal cell in Fourier space. $\Omega_n$ (resp. $\widehat{\Omega}_n$) is the discretized $[N_x \times N_y \times N_z]$ grid of $\Omega$ (resp. $\widehat{\Omega}$) on points $\vec{r}_{\vec n}$ (resp. $\vec{k}_{\vec n}$) defined by
\begin{equation*}
\vec r_{\vec n} = n_x \frac{L_x}{N_x} \, \vec e_x + n_y \frac{L_y}{N_y} \, \vec e_y + n_z \frac{L_z}{N_z} \, \vec e_z \, 
\end{equation*}
with $n_\mu \in \left[0, N_\mu\right]$ and
\begin{equation*}
\vec k_{\vec n} = n_x \frac{1}{L_x} \, \vec e_x + n_y \frac{1}{L_y} \, \vec e_y + n_z \frac{1}{L_z} \, \vec e_z \, 
\end{equation*}
with $n_\mu \in \left[-N_\mu / 2 + 1, N_\mu/ 2 \right]$.

The unitary Fourier transform and its inverse, in ordinary frequency, connecting $\Omega$ and $\widehat{\Omega}$ are given by:
\begin{equation}
\begin{aligned}
\widehat{f}(\vec{k}) &=  \int_\Omega f(\vec{r}) \, e^{-2 \pi i \vec{k} \cdot \vec{r}} d \vec{r}\,, \\
f(\vec{r}) &=  \int_{\widehat{\Omega}} \widehat{f}(\vec{k}) \, e^{2 \pi i \vec{r} \cdot \vec{k}} d \vec{k}\,.
\end{aligned}
\end{equation}

In Fourier space, the solution to the Poisson equation Eq.\eqref{SCF2} is:
\begin{equation}
\begin{aligned}
\widehat{V}^{\text{ee}}(\vec{k}) &= \frac{\widehat{n}(\vec{k}, \mu)}{\pi |\vec{k}|^2}\,,\quad \vec{k} \neq \vec{0} \\
\widehat{V}^{\text{ee}}(\vec{0}) &= 0 \, .
\end{aligned}
\end{equation}

\subsection{Pseudopotential}
\label{pseudo}

The Coulomb potential of the nuclei, $V^{\text{ie}}$, which enters Eq.\eqref{SCF1}, is the sum of the contributions of all $N_i$ nuclei present in the simulation cell $\Omega$ and all their periodic images in the replicas of $\Omega$:
\begin{equation}
V^{\text{ie}}(\vec r) = \sum_{\vec R} \sum_s \sum_{j \in s} V_{s}(\vec r - \vec r_j + \vec R)\, ,
\end{equation}
where $V_{s}(\vec r)$ is the Coulomb potential of the nuclei of species $s$ to which nucleus $j$ belongs, $V_s(\vec r) =  \frac{Z_s}{|\vec r\,|}\,$ and the outer sum over $\vec R$ covers all lattice translation vectors $\vec R = n_x L_x \vec e_x + n_y L_y \vec e_y + n_z L_z \vec e_z$ for all relative integers $(n_x, n_y, n_z)$.

However, $\Omega_n$, the discretized grid of $\Omega$, does not allow for an accurate resolution of the SCF cycle with the Coulomb potential because it is too stiff close to the nucleus. We therefore employ a smoother norm-conserving pseudopotential $\widetilde V_{s}$ \cite{Troullier1991}.
 
The potential $V^{\text{ie}}$ is then obtained in the reciprocal cell $\widehat \Omega_n$ in Fourier space by:
\begin{equation}
\label{eq:iepseudo}
\widehat V^{\text{ie}}(\vec k) = \sum_s \widehat{\widetilde  V_{s}}(\vec k) \, \widehat S_{s}(\vec k)\, ,
\end{equation}
where $\widehat{S_s}$ is the structure factor of species $s$:
\begin{equation}
 \widehat{S_s}(\vec{k}) = \sum_{j \in s} e^{-2 \pi i \vec{k} \cdot \vec{r_j}}  .
\end{equation}

Eq.\eqref{eq:iepseudo} is valid only for $\vec k \neq \vec 0$. Indeed, the pseudopotential behaves like a Coulombic potential at long range, and so $\widehat{\widetilde  V_{s}}(\vec 0)$ diverges. The solution is to simply set $ \widehat{\widetilde  V_{s}}(\vec 0) = 0$, as noted by Payne \cite{Payne1992}: \textit{"[$\ldots$] the total ionic potential at $k=0$ is infinite, so the electron-ion energy is infinite. However, there are similar divergences in the Coulomb energies due to the electron-electron interactions and the ion-ion interactions. The Coulomb $k=0$ contributions to the total energy from the three interactions cancel exactly."}

For each density and temperature, we compute the pseudopotential following the approach proposed by Lambert in Ref. \cite{Lambert2007}. This method is used in OFMD codes such as \textsc{dragon} \cite{MIHAYLOV2024108931} and \textsc{ofmd} \cite{Lambert2006}. First, an electronic density $n(r)$ is obtained within an average-atom model \cite{Feynman1949}. This density is a non-linear response to the stiff Coulomb potential. The norm-conserving method consists of imposing smoother behavior close to the nucleus while conserving the charge within a radius $r_{\text{cut}}$ and the regularity at $r_{\text{cut}}$. The pseudo-density $\tilde{n}(r)$ is then given by:

\begin{equation}
\tilde{n}(r) = 
\begin{cases} 
\exp(a + b r^2 + c r^4), & r < r_{\text{cut}} \\
n(r), & r \geq r_{\text{cut}} 
\end{cases}
\end{equation}

\noindent The parameters $a,b,c$ are determined by the following constraints:
\begin{equation}
\begin{aligned}
& \tilde{n}(r_{\text{cut}}) = n(r_{\text{cut}}) \\
&\left. \frac{\partial \tilde{n}}{\partial r} \right|_{r_{\text{cut}}} = \left. \frac{\partial n}{\partial r} \right|_{r_{\text{cut}}} \\
&\int_{0}^{r_{\text{cut}}} dr \, 4\pi r^2 \tilde{n}(r) = \int_{0}^{r_{\text{cut}}} dr \, 4\pi r^2 n(r)
\end{aligned}
\end{equation}

\noindent Once the pseudo-density is known, the pseudopotential is obtained by inverting Eq.\eqref{SCF1} such that $\tilde{n}$ minimizes the Lagrangian in Eq.\eqref{Lagrangian} when the Coulomb potential of the nuclei is replaced by the pseudopotential:

\begin{eqnarray}
& &\widetilde V_{s}(r) = \mu - k_{\mathrm{B}} T_{\mathrm{e}} I_{1/2}^{-1} \left( \frac{2 \pi^2 \tilde{n}(r)}{(2 k_{\mathrm{B}} T_{\mathrm{e}})^{3/2}}  \right)  ~~~~~~\\
& &- \frac{1}{r} \int_{0}^{r} dx \, 4\pi x^2 \tilde{n}(x) - \int_{r}^{r_{\text{cut}}} dx \, 4\pi x \tilde{n}(x) \nonumber \\
& &- V_{\text{xc}}[\tilde{n}],\nonumber
\end{eqnarray}

\noindent where $I_{1/2}^{-1}$ is the inverse Fermi-Dirac function.

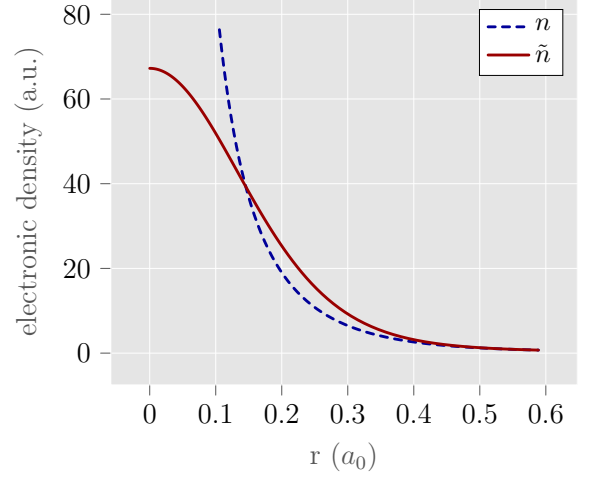
\begin{figure}
\centering
\begin{tikzpicture}[thick, line cap=round, xscale=0.9, yscale=0.9]

\definecolor{dimgray85}{RGB}{85,85,85}
\definecolor{gray119}{RGB}{119,119,119}
\definecolor{gainsboro229}{RGB}{229,229,229}

\def\Lbox{42.509174}

\begin{axis}[
axis background/.style={fill=gainsboro229},
axis line style={white},
tick align=outside,
tick pos=left,
x grid style={white},
xlabel=\textcolor{dimgray85}{r ($a_0$)},
xmajorgrids,
xtick style={color=dimgray85},
y grid style={white},
ylabel=\textcolor{dimgray85}{electronic density (a.u.)},
ymajorgrids,
restrict x to domain=0.:0.6, 
restrict y to domain=0.:80., 
ytick style={color=dimgray85},
legend style={font=\small},
legend cell align=left,
legend pos=north east
]
\addplot [very thick, dashed, blue!60!black] table[x index=1, y index=2] {AA-densities-13.000000.dat};
\addlegendentry{$n$}
\addplot [very thick, red!60!black] table[x index=1, y index=3] {AA-densities-13.000000.dat};
\addlegendentry{$\tilde n$}
\end{axis}

\end{tikzpicture}
\caption{Electronic density $n$ and pseudo-density $\tilde n$ of aluminum at $\rho_0$ at temperature $T = 100$ eV as functions of the radius $r$ in Bohr radius $a_0$.}
\label{fig:regdensity}
\end{figure}

As shown in Fig. \ref{fig:regdensity}, the pseudo-density is less polarized in the vicinity of the nucleus than the exact density. Both the pseudo-density and the pseudopotential deviate from their exact counterparts inside the cutoff radius $r_{\text{cut}}$.
This cutoff radius is chosen as a fraction of the Wigner–Seitz radius $r_{\text{ws}}$. This radius defines the sphere enclosing the average atom and is related to the ionic number density $\rho$ through $$ \frac{4}{3}\pi \rho \, r_{\text{ws}}^{3} = 1\,.$$

Consequently, the pseudopotential is consistent with both the density and the temperature of the simulation. Within the average-atom framework, the pressure depends only on the value of the density $n$ at $r_{\text{ws}}$; it is therefore unaffected by the use of a norm-conserving pseudopotential. The same behavior is expected in OFMD simulations.

In contrast, the total energy depends on the density throughout the entire Wigner–Seitz sphere and is therefore influenced by the pseudo-density. Because the exact density is more localized and more strongly bound near the nucleus, a systematic energy difference arises between the pseudo and exact descriptions. This correction is evaluated within the average-atom model and subsequently applied to the energies obtained from OFMD simulations employing pseudopotentials.

For instance, in the average-atom framework, the energy values obtained for aluminum at $\rho_0$ and 100 eV (corresponding to Fig. \ref{fig:regdensity}) are $-6550$~eV for the Coulombic potential and $-2971$~eV for the pseudopotential.

\subsection{Self-consistent method}
\label{densitycomputation}
Solving Eq.\eqref{SCF1} directly is not possible \textit{a priori} because the chemical potential $\mu$ is unknown, and the potential $V^{\text{ee}}$ depends on $n$ via Eq.\eqref{SCF2}. It is feasible, however, using a self-consistent field (SCF) method combined with Newton's algorithm. This process is described below.

It starts with an input guess: $n_i$, which is $ n_0  \equiv N_e / | \Omega |$ for the first MD timestep (see also Sec.\ref{density:initialization} for subsequent MD timesteps). This guess serves as a source term to solve the Poisson equation, Eq.\eqref{SCF2}, for $V^{\text{ee}}$. Once $V^{\text{ee}}$ is obtained, $V_i$ is the first input potential used as the effective potential $V_{\text{eff}}$ in Eq.\eqref{SCF1}. The latter equation computes an active field $n_a$ with the current value of the chemical potential $\mu$. 

The following equation is then solved to find the value of $\mu$:

\begin{equation}
\label{newtonmu}
\int_\Omega n_a(\vec{r}, \mu) d\vec{r} - N_e = 0.
\end{equation}

To this end, Newton’s method is applied starting from an initial guess $\mu_0$. Given the previously computed $V_i$, it converges within a few iterations (typically fewer than five) to the chemical potential $\mu_a$ and the corresponding electronic density $n_a$.

The electronic density $n_a$ is combined with the input guess $n_i$ (see Sec.\ref{density:mixing}) to generate an SCF iteration density output $n_o$, which is then used as an updated input value $n_i$ for the next SCF iteration. The process is repeated until the difference $||n_i - n_a||_{L^2} < \varepsilon$ is sufficiently small.
 In this method, the density is never negative or null and always complies with the electroneutrality constraint required by Eq.~\eqref{newtonmu}.

 
\subsubsection{Initializing electron density}
\label{density:initialization}
It is possible to speed up the convergence of the SCF loop by setting an initial density close to the solution. During the development of \textsc{Archê}, we observed that from one timestep to the next, the electronic densities were only slightly different. Using this information, our first attempt was to use the last computed electronic density as an initial guess, but this did not improve convergence. What does improve convergence is initializing the density as a function of the nuclei positions, as follows.

We assume that the electron density $n_i(\vec r)$ is the superposition of as many functions $Z_s g_s(\vec r)$ as there are species in the system, centered on each nucleus (which is a realistic approximation in a very dilute gas):

\begin{equation*}
n_i(\vec{r}) = \sum_{\vec R} \sum_s \sum_{j \in s}  Z_s\,g_s(\vec{r} - \vec{r}_j + \vec R)\, ,
\end{equation*}

\noindent where the lattice translation vectors $\vec R = n_x L_x \vec e_x + n_y L_y \vec e_y + n_z L_z \vec e_z$ for all relative integers $(n_x, n_y, n_z)$. The Fourier representation in the reciprocal cell $\widehat \Omega_n$ is given by:
\begin{equation*}
\widehat{n_i}(\vec{k}) = \sum_s Z_s\,\widehat{g_s}(\vec{k})\, \widehat{S}_s(\vec{k}) ,
\end{equation*}
where $\widehat{S_s}$ is the structure factor of species $s$. Now, we assume that all the normalized profiles $g_s(\vec r)$ are equal to a single profile $g(\vec r)$. This assumption yields:
\begin{equation*}
\widehat{n_i}(\vec{k}) =  \widehat{g}(\vec{k})\, \widehat{S}_Z(\vec{k}) ,
\end{equation*}
with the charge-averaged structure factor $\widehat{S}_Z = \sum_s Z_s\,\widehat{S}_s$. Consequently, $\widehat{g} = \widehat{n_i}/\widehat{S}_Z$. 

Therefore, the algorithm consists of computing and storing $\widehat{g^{n}}$ at time $n$ to initialize the density at time $n+1$ using:

\begin{equation}
\widehat{n_i^{n+1}} = \widehat{S_Z^{n+1}} \widehat{g^{n}}
\end{equation}

These additional computations are offset by the resulting speedup in convergence (Sec. \ref{conv:speedup}).

 
\subsubsection{Mixing electron density}
\label{density:mixing}

The mixing algorithm implemented in \textsc{Archê} is an adaptation of the algorithm described in the appendix of More \textit{et al.} \cite{More1988}. Originally designed for an average-atom model, we applied it to OFMD in \textsc{Archê}, where it significantly improved convergence (see Sec. \ref{conv:speedup}). To our knowledge, this has never been done for OFMD simulations. Here, we elaborate on the formulas from More \textit{et al.}'s paper while maintaining the same notation.

The algorithm computes an output density $n_o$ using the active density $n_a$ computed during the SCF cycle (see the beginning of Sec. \ref{densitycomputation}) and the input density $n_i$. Setting $\Delta n = n_i - n_a$, we have $n_o = n_a + w \Delta n$, and this mixture of the two electronic densities serves as the new input density for the next iteration of the SCF loop. In practice, $w$ must often be chosen conservatively --- for example, $n_o = 0.95 n_i + 0.05 n_a$ --- which usually requires many iterations.

More \textit{et al.} utilized all available information. The goal of the algorithm is to select, at each step, an input density that is as close to the converged density as possible. This density extremizes the Lagrangian in Eq.(\ref{Lagrangian}) with respect to variations in the parameter $w$, using densities of the form $n = n_a + w \Delta n$. This density conserves the electroneutrality by construction. Therefore, minimization is achieved when $\frac{d F[n_o(\vec{r})]}{dw} = 0$. It is reasonable to approximate the free energy $F$ using a functional expansion in the small variation $\Delta n$. At second order, this allows us to determine a value of $w$ that minimizes this approximation of $F$:
\begin{eqnarray}
& &F[n(\vec{r})]  \simeq F[n_a(\vec{r})] + w \int_\Omega \frac{\delta F}{\delta n(\vec{r})} \Delta n(\vec{r}) d\vec{r}\nonumber \\
& & + \frac{1}{2} w^2 \int_\Omega \frac{\delta^2 F}{\delta n(\vec{r}) \delta n(\vec{r'})} \Delta n(\vec{r}) \Delta n(\vec{r'}) d\vec{r} d\vec{r'}\nonumber \\
& & + \mathcal{O}(\Delta n^3)\, ,
\end{eqnarray}
\noindent where the functional derivatives are evaluated at $n_a$. This leads to $w = U / (K + U)$, with:
\begin{equation}
U = -\int_\Omega \frac{\delta F}{\delta n(\vec{r})} \Delta n(\vec{r}) d\vec{r},
\end{equation}
\begin{equation}
U + K = \int_\Omega \frac{\delta^2 F}{\delta n(\vec{r}) \delta n(\vec{r'})} \Delta n(\vec{r}) \Delta n(\vec{r'}) d\vec{r} d\vec{r'}.
\end{equation}
\noindent 
The first functional derivative reads:
\begin{equation}
\frac{\delta F}{\delta n(\vec{r})} = \frac{\delta \mathcal{F}_0}{\delta n(\vec{r})} + V_{\text{eff}}.
\end{equation}
For simplicity, we omit the $V_{\text{xc}}$ contribution. Its inclusion is straightforward and does not alter the conclusions. The second functional derivative then reads:
\begin{equation}
\frac{\delta^2 F}{\delta n(\vec{r}) \delta n(\vec{r'})} = \frac{\delta^2 \mathcal{F}_0}{\delta n(\vec{r}) \delta n(\vec{r'})} + \frac{1}{{|\vec{r} - \vec{r}\,'|}}.
\end{equation}
The evaluation of $V_{\text{eff}}$ at $n_a$ is denoted $V_a$. However, $n_a$ is computed with a different potential, $V_i$, which is treated as an external potential. The Euler equation used for the computation of $n_a$ is:
\begin{equation}
\frac{\delta \mathcal{F}_0}{\delta n(\vec{r})} + V_i = \mu,
\end{equation}
leading to:
\begin{eqnarray}
& \dfrac{\delta^2 \mathcal{F}_0}{\delta n(\vec{r}) \delta n(\vec{r'})} &= \dfrac{d\mu}{dn}\, \delta(\vec r - \vec r\,')  \\
& &= \left(\dfrac{dn}{d\mu}\right)^{-1}\, \delta(\vec r - \vec r\,') .\nonumber
\end{eqnarray}
We use the second right-hand side of this equation since $\dfrac{dn}{d\mu}$ is computed during Newton's method to find the value of $\mu$ that ensures electroneutrality. The complete evaluation of the functional derivatives at $n_a$ gives:
\begin{eqnarray}
&\dfrac{\delta F}{\delta n(\vec{r})} &= \mu_a - V_i + V_a \\
& &= \mu_a - \int_\Omega \frac{\Delta n(\vec{r'})}{|\vec{r} - \vec{r'}|}d\vec{r'},\nonumber
\end{eqnarray}
and
\begin{equation}
\frac{\delta^2 F}{\delta n(\vec{r}) \delta n(\vec{r'})} = \left(\dfrac{dn_a}{d\mu_a}\right)^{-1} \delta(\vec r - \vec r\,') + \frac{1}{{|\vec{r} - \vec{r}\,'|}}.
\end{equation}

This derivation led More \textit{et al.} to select this specific value $w = U / (K + U)$ for the mixing parameter, where:
\begin{equation}
U = \int_\Omega \frac{\Delta n(\vec{r})\Delta n(\vec{r'})}{|\vec{r} - \vec{r'}|}d\vec{r}d\vec{r'},
\end{equation}
\begin{equation}
K = \int_\Omega \frac{[\Delta n(\vec{r})]^2}{\partial n_a(\vec{r}) / \partial \mu_a} d\vec{r}.
\end{equation}

In practice, $w$ is computed and updated only after a few SCF iterations, once the chemical potential fluctuations are small. Despite the additional computations, this method pays off by reducing the overall wall-clock time of the simulations.

\subsubsection{Convergence speed up}
\label{conv:speedup}

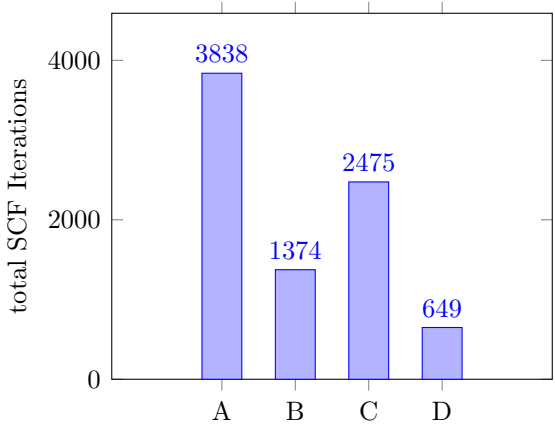
\begin{figure}
\begin{center}
\begin{tikzpicture}[font=\footnotesize]

    \begin{axis}[
    	    scale=0.85,
        ybar,
        symbolic x coords={A, B, C, D},
        xtick=data,
        ymin=0,
        ymax=4590,
        ylabel={total SCF Iterations},
        nodes near coords,
        enlarge x limits=0.5,
        bar width=15pt,
    ]
        \addplot coordinates {(A,3838) (B,1374) (C,2475) (D,649)};
    \end{axis}
\end{tikzpicture}
\end{center}

\caption{Improved convergence of the SCF cycle with respect to a naive setup (A) using a constant mixing parameter and a uniform density initialization. Setup (B) improves the mixing (Sec.\ref{density:mixing}). Setup (C) improves the initialization (Sec. \ref{density:initialization}). The combination of both improvements (D) accelerates convergence by up to a factor of 6 in this typical example of aluminum at 300\,eV and $\rho=4\, \rho_0$, where 128 nuclei are propagated for 101 timesteps on a $256^3$ spatial grid.}
\label{fig:convergence}
\end{figure}

Fig.\ref{fig:convergence} provides a numerical example of the improvements achieved through the new density mixing and initialization algorithms. This example displays the total number of SCF iterations required for a simulation of 101 molecular dynamics timesteps. Algorithm (A) represents the naive setup, which initializes the electron density at the start of the SCF loop with a constant value and mixes the input and active densities in fixed proportions at each iteration to generate an output density. Setup (B) incorporates only the improved mixing algorithm based on More's approach\linebreak(Sec.\ref{density:mixing}). Setup (C) includes only the initialization improvement using a superposition approximation with an average one-center density profile (Sec.\ref{density:initialization}). Finally, version (D) is the complete implementation utilizing both improvements. This example, like many others, demonstrates that when both improvements are combined, the speedup can approach a factor of 6. Furthermore, under many conditions, the mixing algorithm not only accelerates convergence but actively ensures it. This enhancement is highly valuable for both the speed and the robustness of the code --- a combination rare enough to be worth emphasizing.


\subsection{Forces and thermodynamic quantities}
\label{sec:thermo}

Once the total energy of the system is known as a function of nuclear positions and velocities, along with the electron density contribution, the forces acting on the nuclei can be computed as the derivative of the energy with respect to the nuclear positions, according to Eq.\eqref{eq:force}. The pressure is computed as the trace of the stress tensor, whose contributions are obtained either from the free energy via a functional derivative with respect to the deformation tensor \cite{Morante2006} or from the forces using a generalized virial theorem \cite{Nielsen1985A}. 

The different contributions to the total energy are the kinetic energies of the nuclei ($K^\text{i}$) and electrons ($K^\text{e}$), the interaction between nuclei ($U^{\text{ii}}$), the interaction between electrons ($U^{\text{ee}}$), and the interaction between nuclei and electrons ($U^{\text{ie}}$). Therefore, we split the total energy $E$ as follows:
\begin{equation}
 E =  K^\text{i} + K^\text{e} + U^\text{ii} + U^{\text{ie}} + U^{\text{ee}} + \Delta E_\text{AA}.
\end{equation}
A correction $\Delta E_\text{AA}$ is added to compensate for the regularization of the pseudopotential (see Sec.\,\ref{pseudo}).

The forces involve only the contributions $U^\text{ii}$ and $U^\text{ie}$, which depend explicitly on the nuclear positions. As previously stated:
\begin{eqnarray}
&\vec{F}_j(t) &= \vec{F}_j^{ii}(t) + \vec{F}_j^{ie}(t) \\
& &= -\nabla_{\vec{r}_j}(U^{ii}(R(t)) + U^{ie}(R(t))).\nonumber
\end{eqnarray}

The pressure can be derived from the trace of the so-called average virial stress tensor (also known as the pressure tensor; see \cite{ZHOU2003, Bartolotti1980}):
\begin{equation}
P = - \frac{1}{3} \text{tr}(\boldsymbol{\Sigma}).
\end{equation}
There are as many contributions to the stress tensor as there are to the energy (excluding the correction $\Delta E_\text{AA}$, which arises from comparing the Coulombic and pseudopotentials within the average atom model):
\begin{equation}
\boldsymbol{\Sigma} =  \boldsymbol{\sigma^i} + \boldsymbol{\sigma^e} + \boldsymbol{\sigma^{ii}} + \boldsymbol{\sigma^{ee}} + \boldsymbol{\sigma^{ie}}.
\end{equation}

\subsubsection{Kinetic energy of nuclei}
The kinetic contributions follow from the principles of statistical mechanics \cite{Morante2006}. The kinetic energy is given by:
\begin{equation}
\label{def::K_i}
K^i = \frac{1}{2}\sum^{N_i}_i m_i \boldsymbol{v}^2_i\, ,
\end{equation}
and its contribution to the stress tensor reads:
\begin{equation}
\label{def::sigma_i}
\sigma^i_{\mu \nu} = -\frac{2}{|\Omega|}  \sum^{N_i}_i \frac{m_i v_{\mu} v_{\nu}}{2}\, .
\end{equation}

\subsubsection{Kinetic energy of electrons}

The TF kinetic energy is given by \cite{Nikiforov2005, Lambert2007, Starrett2017}:
\begin{equation}
K_e = k_B T_e  \int \mathrm{d}\vec{r}\,  c_{\text{TF}} \, \mathrm{I}_{3/2}[\eta(\vec{r})] 
\end{equation}
\noindent with $c_{\text{TF}} = \frac{\sqrt{2}}{\pi^2} (k_B T_e)^{3/2} $ and $$\eta(\vec{r}) = \left[\mu - V_{\text{eff}}(\vec{r}) \right] / (k_B T_e).$$

Since the non-interacting free energy density $f_0$, defined by $\mathcal{F}_0 = \int \mathrm{d}\vec{r}\ f_0(n)$, is a local functional of the density $n$ in the TF model, the stress is diagonal and reads:
\begin{eqnarray}
&\sigma^e_{\mu \nu} &= \frac{1}{|\Omega|}\int \mathrm{d}\vec{r}\,\left[f_0[n] - n \frac{\partial f_0}{\partial n} \right] \delta_{\mu \nu} \nonumber\\
& &= -\frac{2 K_e}{3 |\Omega|}\, \delta_{\mu \nu} .
\end{eqnarray}
Corrections to the TF model incorporate the density gradient through semi-local functionals of the von Weizsäcker type \cite{Mi2023, Karasiev_2025, Luo2020}, as well as non-local functionals involving density values evaluated at two different spatial positions \cite{Mi2023, Karasiev_2025, RiosVargas2024, Bhattacharjee2024}. Expressions for the corresponding contributions to the kinetic energy and stress tensor can be found in the supplemental material of Ref.\cite{Sjostrom2014}.

\subsubsection{Interaction energy between nuclei}

The interaction between nuclei requires summing the long-range Coulomb interactions between the particles in the simulation box and all their infinite periodic images. This is evaluated using Ewald summation, and is therefore split into three components: short-range (particle-particle), long-range (particle-mesh), and a neutralizing jellium background. The separation between the long- and short-range components is controlled by the Ewald parameter $\alpha$ \cite{Kittel2004, Toukmaji1996}.
\begin{equation}
U^\text{ii} = U^\text{pp} + U^\text{pm} + U^\text{je} 
\end{equation}

We selected a value for $\alpha$ that allows the short-range interaction to be evaluated using the minimum image convention \cite{MecaStatsTuckerman}.

The short-range energy $U^\text{pp}$ reads:
\begin{equation}
U^\text{pp} = \sum^{N_i}_{j = 1} \sum^{N_i}_{k > j} Z_j Z_k \frac{\operatorname{erfc} (\alpha |\vec{r}_{jk}|)}{|\vec{r}_{jk}|}\,.
\end{equation}
The long-range energy $U^{pm}$ reads:
\begin{eqnarray}
\label{eq:ewald:longrange}
&U^\text{pm} &= \frac{1}{2 |\Omega|} \sum_{\vec{k}\,\in\,\widehat{\Omega}_n,\,\vec{k} \neq 0} \widehat{V}_{\alpha}(\vec k) | \widehat{S}_Z(\vec k)  |^2 \nonumber\\
& &- \frac{\alpha}{\sqrt{\pi}} \sum^{N_i}_{j = 1} Z_j^2\,,
\end{eqnarray}
with:
\begin{equation}
\widehat{V}_{\alpha}(\vec k) = \frac{1}{\pi} \frac{e^{- ({\pi k}/{\alpha})^2}}{k^2}  .
\end{equation}

The value $\vec k = 0$ is excluded from the summation because the system is neutralized by a jellium of opposite charge $Q = \sum_j^{N_i} Z_j$, as explained in Appendix F of Ref.\cite{Martin2020}. The second term on the right-hand side of Eq.~\eqref{eq:ewald:longrange} corrects for the self-interaction energy that was artificially introduced in the first one.

The energy contribution due to the jellium is given by: 
\begin{equation}
E^\text{je} = - \frac{\pi}{2 \alpha^2 |\Omega|} Q^2 .
\end{equation}

The jellium does not contribute to the forces, which are split into short-range (pp) and long-range (pm) components as follows:
\begin{equation}
\vec{F}_j^{ii}(t) = \vec{F}_j^\text{pp} + \vec{F}_j^\text{pm} ,
\end{equation}

\begin{eqnarray}
& &\vec{F}^\text{pp}_j = - \sum^{N_i}_{k \neq j} Z_j Z_k  \\
& & \times \left( \frac{\operatorname{erfc} (\alpha |\vec{r}_{jk}|)}{|\vec{r}_{jk}|} + \frac{2 \alpha}{\sqrt{\pi}} \operatorname{exp}(-\alpha^2 |\vec{r}_{jk}|^2)  \right)\frac{\vec{r}_{jk}}{ |\vec{r}_{jk}|^2} ,\nonumber
\end{eqnarray}
and
\begin{eqnarray}
&\vec{F}^\text{pm}_j  = &Z_j\,\dfrac{2 \pi}{|\Omega|}~ \sum_{\vec{k}\,\in\,\widehat{\Omega}_n,\,\vec{k} \neq 0} \widehat{V}_{\alpha}(\vec k) \\
& &\times \,\text{Im}\left(  e^{i 2 \pi \vec{k} \cdot \vec{r}_j} \widehat{S}_Z(\vec k) \right) \vec{k} .\nonumber
\end{eqnarray}

The jellium contributes to the stress tensor, which is also split into short-range (pp) and long-range (pm) components:
\begin{equation}
\boldsymbol{\sigma^\text{ii}} = \boldsymbol{\sigma^\text{pp}}  + \boldsymbol{\sigma^\text{pm}} + \boldsymbol{\sigma^\text{je}} ,
\end{equation}
with
\begin{eqnarray}
& &\sigma^\text{pp}_{\mu \nu} = -\frac{1}{|\Omega|} \sum^{N_i}_{i = 1} \sum^{N_i}_{j > i} Z_i Z_j \\
& &\times\left( \frac{\operatorname{erfc} (\alpha |\vec{r}_{ij}|)}{|\vec{r}_{ij}|} + \frac{2 \alpha}{\sqrt{\pi}} \operatorname{exp}(-\alpha^2 |\vec{r}_{ij}|^2)  \right) \frac{r_{ij}^\mu r_{ij}^\nu}{ |\vec{r}_{ij}|^2} ,\nonumber
\end{eqnarray}
\begin{eqnarray}
&\sigma^\text{pm}_{\mu \nu} &= \frac{1}{2 |\Omega|^2} \sum_{\vec{k}\,\in\,\widehat{\Omega}_n,\,\vec{k} \neq 0} \widehat{V}_{\alpha}(\vec k)\\
& & \times \left| \widehat{S}_Z(\vec k) \right|^2\left[ 2 k_\mu k_\nu \left( \frac{\pi^2}{ \alpha^2} + \frac{1}{k^2} \right) - \delta_{\mu \nu}\right] ,\nonumber
\end{eqnarray}
and
\begin{equation}
\sigma^\text{je}_{\mu \nu} = \frac{\pi}{2 \alpha^2 |\Omega|^2} Q^2\, \delta_{\mu \nu}\ .
\end{equation}


\subsubsection{Interaction energy between electrons}

The electrostatic energy can be computed in several equivalent ways:

\begin{eqnarray}
&U^\text{ee} &= \frac{1}{2 } \int_{\Omega}  n (\vec{r}) \, V^\text{ee} (\vec{r}) \, d\vec{r}\\
& &= \frac{1}{2 } \int_{\widehat \Omega}  \widehat n (\vec{k}) \, \overline{\widehat V^\text{ee}} (\vec{k}) \, d\vec{k} \,,\nonumber
\end{eqnarray}
with their discretized versions:
\begin{eqnarray}
&U^\text{ee} &= \frac{|\Omega|}{2N} \sum_{\vec{r}\,\in\,{\Omega}_n}\,  n (\vec{r}) \,V^{\text{ee}} (\vec{r}) \\
& &= \frac{1}{2 |\Omega|} \sum_{\vec{k}\,\in\,\widehat{\Omega}_n,\,\vec{k} \neq 0}  \widehat{n} (\vec{k}) \,\overline{\widehat V^\text{ee}} (\vec{k}) \,. \nonumber
\end{eqnarray}

The tensor formula is derived from Appendix G of Ref.\cite{Martin2020}:

\begin{equation}
\sigma^\text{ee}_{\mu \nu} = \frac{1}{2 |\Omega|^2} \sum_{\vec{k}\,\in\,\widehat{\Omega}_n,\,\vec{k} \neq 0} \widehat{n}(\vec{k}) \overline{ \widehat{V}^\text{ee} }(\vec{k}) \left[ 2 \frac{k_\mu k_\nu }{k^2}  - \delta_{\mu \nu}\right]\, .
\end{equation}

Quantum corrections due to exchange and correlation are not considered here. They can take the form of local density functionals or semi-local functionals that depend on the density gradient \cite{Karasiev_2025}.

\subsubsection{Interaction energy between nuclei and electrons}

The general formula incorporating $V^{\text{ie}}$, the complete pseudopotential defined in Eq.\eqref{eq:iepseudo}, is:
\begin{eqnarray}
&E^\text{ie} &= \int_\Omega n(\vec{r}) V^{\text{ie}} (\vec r) \,d\vec{r} \\
& &-\langle n \rangle \sum_{s} \int_0^{\infty} N_s \left( \widetilde{V}_s(r) - \frac{Z_s}{r} \right) 4 \pi r^2 dr.\nonumber
\end{eqnarray}
where $\langle n \rangle = Q / |\Omega|$ and $N_s$ is the total number of ions of species $s$. The second term on the right-hand side accounts for the non-Coulomb part of the pseudopotential at $k=0$ (see Sec. II.D.3 of Ref.\cite{Payne1992}).

For each nucleus $j$ of species $s$, the total electron force acting upon it involves the pseudopotential $\widetilde{V}_s$, according to (see Ref. \cite{Martin2020} for the derivation and \cite{Profess2008} for an example implementation in the \textsc{profess} code):
\begin{eqnarray}
 \label{Fie}
 &\vec{F}^{ie}_{j,s} &= -\nabla_{\vec{r}_j}(U^{ie}(R(t)))  \\
 & &= 2\pi \sum_{\vec{k}\,\in\,\widehat{\Omega}_n} \widehat{\widetilde{V}}_s(\vec{k})\operatorname{Im}\left(\,\overline{\widehat{n}(\vec{k})} e^{-2\pi i \vec{k}\cdot \vec{r_j}}\right) \vec{k}\,.\nonumber
\end{eqnarray}

Accordingly, the stress is given by \cite{Martin2020}:
\begin{eqnarray}
& &\sigma^\text{ie}_{\mu \nu} =  -\frac{E^\text{ie} }{|\Omega|}  \delta_{\mu \nu}\\
& &-\frac{1}{|\Omega|^2} \sum_{\vec{k} \neq 0} \widehat{n}(\vec{k}) \frac{k_\mu k_\nu}{|\vec k |}  \sum_s \sum_{j \in s} \frac{\partial \widehat{\widetilde{V_s}} (|\vec k|)}{\partial|\vec k|} \, e^{ 2 \pi i \vec{k} \cdot \vec{r}_j} \nonumber
\end{eqnarray}


\begin{figure*}[h!]
\centering
\begin{tikzpicture}[node distance=2cm, scale=0.65, every node/.style={scale=0.65}]

\node (in1) [io] {Input: initial microstate at $t = t_0$};
\node (pro1) [process, below of=in1,yshift=0.5cm] {Compute positions};

\node (pro2) [process, below of=pro1,yshift=0.5cm] {Compute $F_{ii}$

\textit{(Ewald)}};

\node (fie1) [io, below of=pro2, yshift=0.5cm] {initial $n_{e}$, $V_{ie}$};
\node (fie2) [process, right of=fie1,xshift=2cm] {Compute~$V_{e}$ $\Delta V_{ee} = - 4 \pi n$};
\node (new1) [io, right of=fie2, xshift=2.5cm] {initial $\mu$, $n_{e}$, $V_{ie} + V_{ee}$};

\node (new2) [process, below of=new1, yshift=0.5cm] {\textit{Fermi-Dirac Integrals}};
\node (new3) [decision, below of=new2, yshift=-0.2cm] {$|\delta \mu| \leq \varepsilon$};
\node (new4) [process, right of=new3, xshift=1.5cm] {$\mu^{n+1} = \mu^{n} + \delta \mu$};

\node (decne) [decision, left of=new3, xshift=-2.5cm] {$|\delta n| \leq \varepsilon$};

\node (fie4) [process, below of=fie2,yshift=0.5cm] {$n^{n+1}_e = n^{n} + \delta n$};

\node (pro3) [process, left of=decne,xshift=-2.0cm] {Compute $F_{ie}$};
\node (pro4) [process, below of=pro3,yshift=0.5cm] {Compute velocities};
\node (dec1) [decision, below of=pro4] {$t  \leq t_f$};

\node (pro2b) [process, left of=dec1, xshift=-2cm] {Create new microstate};
\node (out1) [io, right of=dec1, xshift=3.5cm] {Output:  thermodynamics quantities};
\node (stop) [startstop, right of=out1, xshift=4cm] {Stop};

\draw [arrow] (in1) -- (pro1);
\draw [arrow] (pro1) -- (pro2);
\draw [arrow] (pro2) -- (fie1);
\draw [arrow] (fie1) -- (fie2);
\draw [arrow] (fie2) -- (new1);

\draw [arrow] (new1) -- (new2);
\draw [arrow] (new2) -- (new3);

\draw [arrow] (new3) -- node[anchor=south] {no} (new4);
\draw [arrow] (new4) |- (new2);

\draw [arrow] (pro3) -- (pro4);
\draw [arrow] (new3) -- node[anchor=south] {yes} (decne);
\draw [arrow] (decne) -- node[anchor=south] {yes} (pro3);
\draw [arrow] (decne) -- node[anchor=west] {no} (fie4);

\draw [arrow] (fie4) -- (fie2);

\draw [arrow] (pro4) -- (dec1);

\draw [arrow] (dec1) -- node[anchor=south] {no} (out1);
\draw [arrow] (dec1) -- node[anchor=south] {~yes} (pro2b);
\draw [arrow] (pro2b) |- (pro1);
\draw [arrow] (out1) -- (stop);

\draw [thick, loosely dashed, black!50] (-2.1,-0.7) rectangle (14.1,-10.5);
\node (SCF) at (12.5, -1.2) {\textit{MD loop} $\times 3000$};
\draw [thick, dotted, black!50] (1.9,-1.6) rectangle (13.9,-9.6);
\node (SCF) at (12.5, -2.2) {\textit{SCF loop} $\times 30$};
\draw [thick, densely dashed, black!50] (5.8,-2.7) rectangle (13.7,-9.4);
\node (NEW) at (11.4, -3.2) {\textit{Electroneutrality loop} $\times 3$};

\end{tikzpicture}
\caption{Flow chart of the core simulation, where the main loop consists of a single MD iteration. Each iteration requires computing the electronic density via a self-consistent field (SCF) loop given by equations (\ref{SCF1}) and (\ref{SCF2}). Newton's algorithm guarantees that the plasma remains neutral in the simulation box, subject to the constraint in equation (\ref{defN}), and is performed for each SCF iteration. The number of iterations shown for each loop is illustrative and reflects typical orders of magnitude.}
\label{fig:flowchart}
\end{figure*}
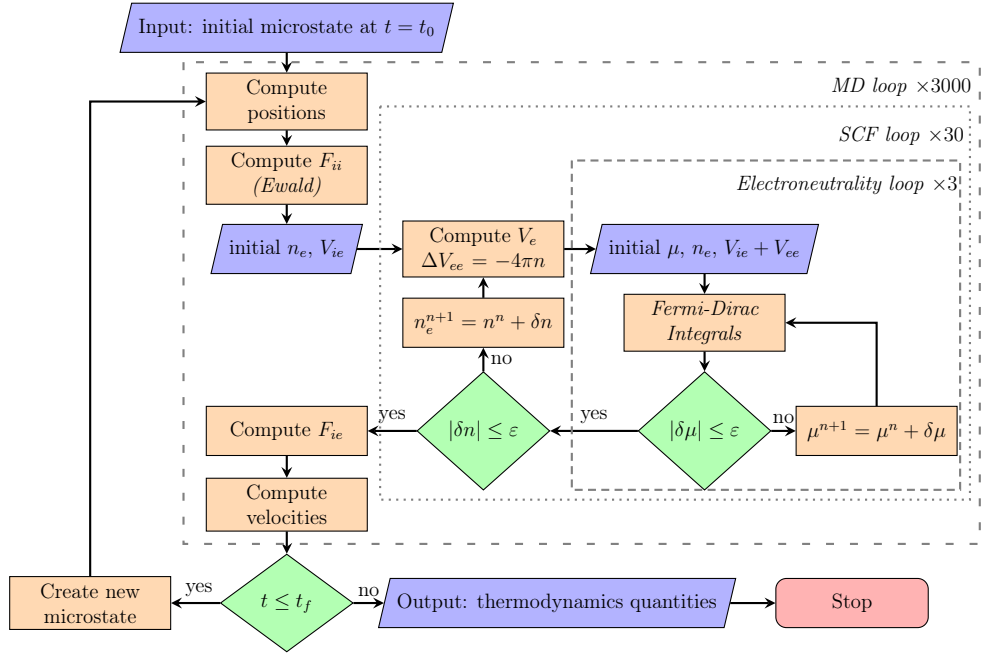
\section{\textsc{Archê} implementation}
\label{implementation}

\begin{table}[b!]
\begin{center}
\caption{Key statistics of the \textsc{Archê} codebase}
\label{CodeNumber}
\begin{tabular}{c|c}

\begin{tabular}{l||c|c}
Language & Files & Lines \\
\hline
CMake & 58 & 1029 \\
Python & 28 & 4887 \\
C/C++ & 224 & 29430 \\
CUDA & 22 & 4671\\
\end{tabular}
&
\begin{tabular}{rl}
~ & Tests \\
\hline
22 & Unit  \\
10 & Integration \\
 &  \\
 & \\
\end{tabular}
\\
\end{tabular}

\end{center}
\end{table}

\textsc{Archê} is the successor to the Fortran code \textsc{ofmd} developed by Lambert \cite{Lambert2007}. The most computationally demanding parts are written in C++ and CUDA, while Python is used for pre- and post-processing. It is parallelized using MPI and relies on the HeFFTe fast Fourier transform library \cite{HEFFTEfirst}. It can run on both CPU and GPU architectures, achieving an average ten-fold speedup on a single GPU compared to 256 CPU cores. It is interfaced with the \textsc{libxc} library for exchange-correlation functionals \cite{LibXC2018} at zero temperature and includes its own implementation of the finite-temperature exchange-correlation functional KSDT \cite{xcKSDT}, translated from its Fortran implementation in the \textsc{Abinit} code \cite{AbinitKSDT2025}. Table \ref{CodeNumber} provides key statistics regarding the codebase.
 
The flowchart in Fig. \ref{fig:flowchart} summarizes the overall architecture of the \textsc{Archê} code. Calculation loops are represented by three rectangles enclosing their respective routines. It illustrates that at each MD timestep, the self-consistent solution is computed within a nested SCF loop, which itself contains another nested loop: the Newton's algorithm used to determine the chemical potential. SCF iterations account for over 95\% of the total computation time on CPUs. Consequently, the number of SCF iterations is the primary factor determining execution time. For this reason, our initial development efforts focused on reducing this number (see Sections \ref{density:initialization} and \ref{density:mixing}).

The SCF loop is implemented in two versions. The first is written in standard C++ for CPU architectures, while the second is implemented in C++/CUDA for GPUs. FFT calls are handled through a wrapper that dispatches them to the appropriate FFT library depending on the architecture (CPU or GPU). Both versions are parallelized using MPI. Only the electronic density calculation is parallelized via domain decomposition in both real and reciprocal spaces, primarily to leverage parallel FFTs. Communication costs are the main bottleneck for the scalability of the Fourier transforms, and consequently for the entire computation, on both CPU and GPU architectures. This effect is even more pronounced on GPUs.

\textsc{Archê} was validated through term-by-term comparisons with Lambert’s \textsc{ofmd} code. Development was guided by these comparisons, and updates were accepted only when relative differences were below $10^{-5}$ for all force components. After successive iterations, excellent agreement between \textsc{ofmd} and \textsc{Archê} was achieved, and unit tests were implemented to prevent any future regressions in the results.

\section{Validation}
\label{validation}
 
\begin{figure}
\begin{center}
\begin{tikzpicture}[xscale=0.9, yscale=0.9]
\definecolor{blue040255}{RGB}{0,40,255}
\definecolor{darkorange2551480}{RGB}{255,148,0}
\definecolor{deepskyblue0212255}{RGB}{0,212,255}
\definecolor{dimgray85}{RGB}{85,85,85}
\definecolor{dodgerblue0128255}{RGB}{0,128,255}
\definecolor{gainsboro229}{RGB}{229,229,229}
\definecolor{gold2552290}{RGB}{255,229,0}
\definecolor{greenyellow19225554}{RGB}{192,255,54}
\definecolor{lightgreen124255121}{RGB}{124,255,121}
\definecolor{maroon12700}{RGB}{127,0,0}
\definecolor{mediumblue00222}{RGB}{0,0,222}
\definecolor{navy00127}{RGB}{0,0,127}
\definecolor{orangered255700}{RGB}{255,70,0}
\definecolor{red22200}{RGB}{222,0,0}
\definecolor{turquoise54255192}{RGB}{54,255,192}

\begin{axis}[
axis background/.style={fill=gainsboro229},
axis line style={white},
tick align=outside,
tick pos=left,
x grid style={white},
xlabel=\textcolor{dimgray85}{time (a.u)},
xmajorgrids,
xtick style={color=dimgray85},
y grid style={white},
ylabel=\textcolor{dimgray85}{Pressure (a.u)},
ymajorgrids,
ytick style={color=dimgray85},
]

\draw[blue!20!black,fill=violet!70!black!20, rounded corners=10,thick,shift={(30,-10)},scale=1.5]
    (0,0) rectangle (85,110)
  node[above, xshift=-2.4cm] {$\langle P \rangle$};
       
\addplot [semithick, mydarkred!60!yellow!50]
table [x index=0,y index=3]{thermodynamics-thermalization.dat};

\addplot [only marks, mydarkred!50!red!70, mark size=1, mark=x]
table [x index=0,y index=3]{thermodynamics-thermalization.dat};

\end{axis}

\end{tikzpicture}
\caption{Example of a pressure trajectory during a dynamic molecular simulation. The average is computed after thermalization has occurred. The data acquisition phase corresponds to the colored area (roughly 80\% of the total duration).}
\label{fig:pressurehistory}
\end{center}
\end{figure}
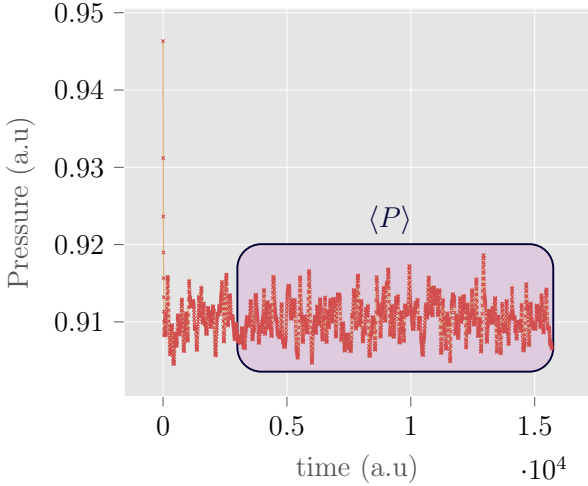

\textsc{Archê} was validated by comparing its internal energy and pressure results for aluminum with those obtained using the \textsc{Extended KSDFT}\linebreak model implemented in \textsc{Abinit} \cite{BlanchetAluminum2022}. This reference model, hereafter referred to as \textsc{Abinit Extended}, explicitly accounts for electronic orbitals.

Simulations were run on GPU using a discrete 3D spatial grid of $256^3$ points and the Thomas-Fermi functional. The timestep $\Delta t$ was chosen to be less than one-tenth of the smaller of the inverse plasma frequency $\tau_{\mathrm{p}}$ and the ratio of the system length $L$ to the thermal velocity $v_{\mathrm{th}}$:
$$\Delta t < 0.1\min\left(\tau_{\mathrm{p}}, L/v_{\mathrm{th}}\right).$$ 
     
To extract an equation of state from a simulation, the contributions to the internal energy and pressure are averaged over time. As shown in Fig.\ref{fig:pressurehistory}, this must be done after a thermalization phase, which, in our simulations, represents around 20\% of the total duration of 3000 timesteps. Several system sizes were tested $N_i \in \{32, 108,$ $128, 256\}$, and the simulation yielding the smallest variance was selected. Variances were computed using non-correlated values, following \cite{Jaupart2021}. In the present study, relative errors are less than $1\%$.

In all comparison figures, \textsc{Archê} results are shown as solid lines with '+' markers, while those obtained with \textsc{Abinit Extended} are represented by 'o' circles. Agreement is considered good when the '+' is centered within the circle, corresponding to a relative error of less than 5\%. In some configurations (e.g., at low temperatures), \textsc{Archê} cannot match the reference, and the discrepancies can become quite significant (such as an order of magnitude for energies). Because the present study relies solely on a Thomas-Fermi model, agreement is naturally good at high temperatures but deteriorates at lower temperatures, as expected.

\begin{figure}[t!]
\centering
\input{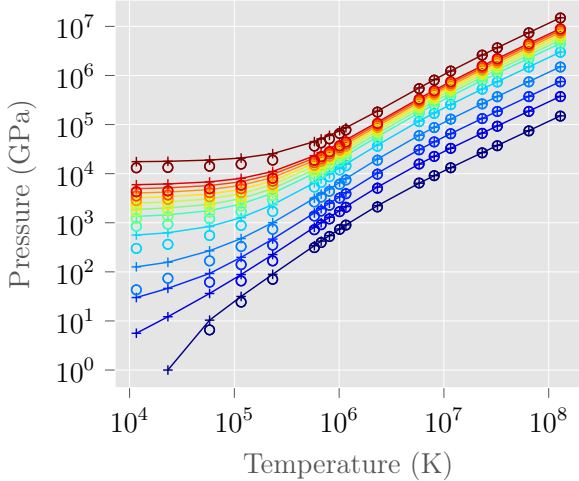} \\
\caption{Pressure along aluminum isochores. \textsc{Archê} results are shown as solid lines with '+' markers; \textsc{Abinit Extended} results are shown with circles. Compression decreases from top $(10\rho_0)$ to bottom $(0.1\rho_0)$ (see Fig.\ref{fig:alu:isotherme} for density values).}
\label{fig:alu:isochore}
\end{figure}

\begin{figure}[t!]
\centering
\input{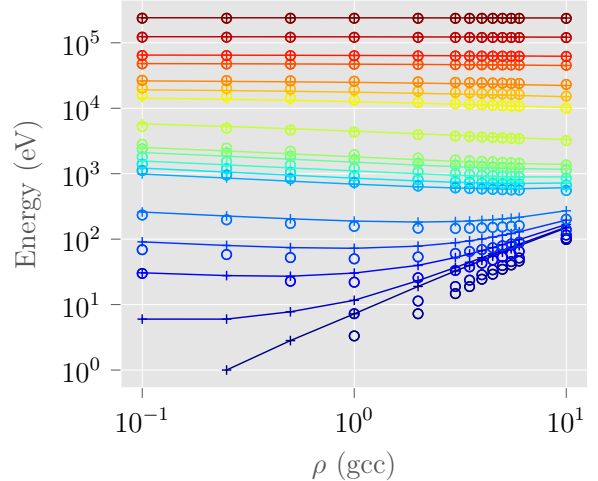} \\
\caption{Energies along aluminum isotherms. \textsc{Archê} results are shown as solid lines with '+' markers; \textsc{Abinit Extended} results are shown with circles. Temperature decreases from top ($10^8$ K) to bottom ($10^4$ K) (see Fig.\ref{fig:alu:isochore} for temperature values).} 
\label{fig:alu:isotherme}
\end{figure}

Fig. \ref{fig:alu:isochore} compares the pressures obtained by \textsc{Archê} and \textsc{Abinit Extended}. Fig. \ref{fig:alu:isotherme} compares the energies along the isotherms with the \textsc{Abinit Extended} results. Since energy is defined only up to an additive constant, the \textsc{Archê} energies had to be shifted. We applied a uniform correction equal to the difference between the lowest energy value obtained with \textsc{Abinit Extended} and the corresponding \textsc{Archê} value. This same correction was added to all \textsc{Archê} energies. Visible differences appear at low temperatures, but for temperatures above 10 eV ($\sim 10^5$ K), the energies obtained are close (within 1\%).


In addition to aluminum, boron and iron were also compared with \textsc{Abinit Extended} \cite{BlanchetBoron2022,BlanchetIron2025}, yielding similar results: good agreement at high temperatures, but reduced accuracy at lower temperatures. Incorporating a gradient correction and, most notably, the KSDT exchange-correlation functional significantly improves agreement at low temperatures. A more in-depth analysis of these results will be presented in a future paper.

\begin{figure*}[t!]
\centering
\begin{tikzpicture}[node distance=2cm, scale=0.65, every node/.style={scale=0.65}]

\node (in1) [io] {Input: initial microstate at $t = t_0$};
\node (pro1) [process, below of=in1,yshift=0.5cm] {$\mathcal{O}(N_i)$};

\node (pro2) [highcost, below of=pro1,yshift=0.5cm] {$\mathcal{O}(N_i^2)$};

\node (fie1) [iohigh, below of=pro2, yshift=0.5cm] {$\mathcal{O}(N_i N_g)\times 2$};
\node (fie2) [highcost, right of=fie1,xshift=2cm] {$\mathcal{O}(N_g \log N_g)$};

\node (new1) [io, right of=fie2, xshift=2.6cm] {initial $\mu$, $n$, $V_{ie} + V_{ee}$};

\node (new2) [process, below of=new1, yshift=0.5cm] {$\mathcal{O}(N_g)$};
\node (new3) [decision, below of=new2, yshift=-0.2cm] {$|\delta \mu| \leq \varepsilon$};
\node (new4) [nocost, right of=new3, xshift=1.5cm] {$\mu^{n+1} = \mu^{n} + \delta \mu$};

\node (decne) [decision, left of=new3, xshift=-2.6cm] {$|\delta n| \leq \varepsilon$};

\node (fie4) [highcost, below of=fie2,yshift=0.5cm] {$\mathcal{O}(N_g \log N_g)$};

\node (pro3) [highcost, left of=decne,xshift=-2.0cm] {$\mathcal{O}(N_i N_g)$};
\node (pro4) [process, below of=pro3,yshift=0.5cm] {$\mathcal{O}(N_i)$};
\node (dec1) [decision, below of=pro4] {$t  \leq t_f$};

\node (pro2b) [process, left of=dec1, xshift=-2cm] {Create new microstate};
\node (out1) [io, right of=dec1, xshift=3.5cm] {Output:  thermodynamics quantities};
\node (stop) [startstop, right of=out1, xshift=4cm] {Stop};

\draw [arrow] (in1) -- (pro1);
\draw [arrow] (pro1) -- (pro2);
\draw [arrow] (pro2) -- (fie1);
\draw [arrow] (fie1) -- (fie2);
\draw [arrow] (fie2) -- (new1);

\draw [arrow] (new1) -- (new2);
\draw [arrow] (new2) -- (new3);

\draw [arrow] (new3) -- node[anchor=south] {no} (new4);
\draw [arrow] (new4) |- (new2);

\draw [arrow] (pro3) -- (pro4);
\draw [arrow] (new3) -- node[anchor=south] {yes} (decne);
\draw [arrow] (decne) -- node[anchor=south] {yes} (pro3);
\draw [arrow] (decne) -- node[anchor=west] {no} (fie4);

\draw [arrow] (fie4) -- (fie2);

\draw [arrow] (pro4) -- (dec1);

\draw [arrow] (dec1) -- node[anchor=south] {no} (out1);
\draw [arrow] (dec1) -- node[anchor=south] {~yes} (pro2b);
\draw [arrow] (pro2b) |- (pro1);
\draw [arrow] (out1) -- (stop);

\draw [thick, loosely dashed, black!50] (-2.1,-0.7) rectangle (14.1,-10.5);
\node (SCF) at (12.5, -1.2) {\textit{MD loop} $\times 3000$};
\draw [thick, dotted, black!50] (1.9,-1.6) rectangle (13.9,-9.6);
\node (SCF) at (12.5, -2.2) {\textit{SCF loop} $\times 30$};
\draw [thick, densely dashed, black!50] (5.8,-2.7) rectangle (13.7,-9.4);
\node (NEW) at (11.4, -3.2) {\textit{Electroneutrality loop} $\times 3$};

\end{tikzpicture}
\caption{Flowchart detailing the algorithmic complexities of the main routines. $N_i$ is the number of nuclei, and $N_g$ is the number of grid points.}
\label{fig:flowcomplexity}
\end{figure*}
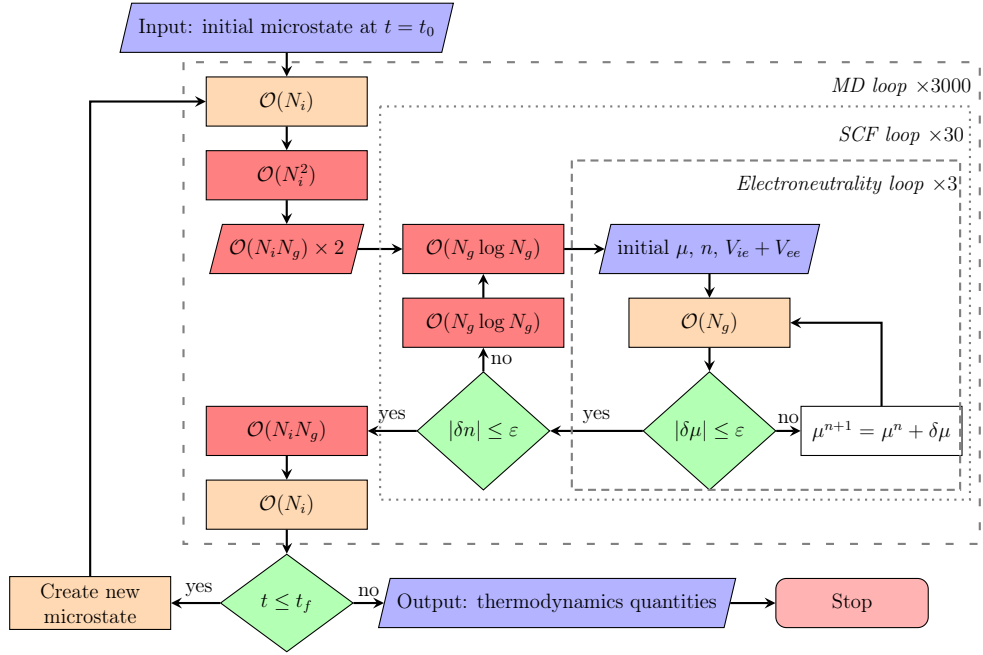

\section{Benchmarks}
\label{benchmark}

Computations were performed on a TGCC supercomputer \cite{TGCC} located in Bruyères-le-Châtel, France. Each node is equipped with two AMD Milan EPYC 7763 processors (64 cores each),\linebreak 256~GB of memory, and four NVIDIA A100 \linebreak(80~GB) GPU accelerators. 
 

\subsection{CPU benchmarks}

The following figures illustrate how the total simulation time varies as a function of specific physical parameters (temperature, density, and number of nuclei $N_i$) as well as numerical parameters (number of grid points $N_g$ and number of CPU cores). These benchmarks provide insights into the algorithmic complexity and scalability of the code. The algorithmic complexity of the various code components is presented in Fig.\ref{fig:flowcomplexity}.
The most computationally intensive component is the SCF loop, which includes the FFT resolution of the Poisson equation that has a complexity of $\mathcal{O}(N_g \log N_g)$ and the evaluation of the Fermi-Dirac integrals across the $N_g$ grid points. Both procedures are executed during each SCF iteration. Consequently, the total number of SCF iterations significantly contributes to the overall computation time. Outside the SCF loop, the other time-consuming components are the calculations of the ion-ion ($ii$) and ion-electron ($ie$) forces, which have complexities of $\mathcal{O}(N_i^2)$ and $\mathcal{O}(N_i N_g)$, respectively.

\subsubsection{Scalability}

\begin{figure} [b!]
    \centering
\begin{tikzpicture}[xscale=0.8, yscale=0.8]

\definecolor{chocolate2267451}{RGB}{226,74,51}
\definecolor{dimgray85}{RGB}{85,85,85}
\definecolor{gainsboro229}{RGB}{229,229,229}
\definecolor{gray119}{RGB}{119,119,119}
\definecolor{lightpink255181184}{RGB}{255,181,184}
\definecolor{mediumpurple152142213}{RGB}{152,142,213}
\definecolor{sandybrown25119394}{RGB}{251,193,94}
\definecolor{steelblue52138189}{RGB}{52,138,189}
\definecolor{yellowgreen14218666}{RGB}{142,186,66}

\begin{axis}[
axis background/.style={fill=gainsboro229},
axis line style={white},
grid=major,
ymin=500,
ymax=40000,
log basis x={2},
log basis y={10},
xmode=log,
ymode=log,
tick align=outside,
tick pos=left,
x grid style={white},
xlabel=\textcolor{dimgray85}{CPU core number},
xtick style={color=dimgray85},
y grid style={white},
ylabel=\textcolor{dimgray85}{Total Time(s)},
ytick style={color=dimgray85}
]

\addplot [line width=1.32pt, violet!70]
table {%
32 18882
64 9563
128 5170
256 2736
512 1636
1024 1130
2048 926
};
\addlegendentry{5000 eV, $N_i = 108$}

\addplot [line width=1.32pt, dashed, steelblue52138189]
table {%
32 15330
64 7848
128 4147
256 2202
512 1305
1024 873
2048 660
};
\addlegendentry{1000 eV, $N_i = 108$}

\addplot [line width=1.32pt, dotted, red!60]
table {%
32 30070
64 15035
128 7912
256 4131
512 2369
1024 1420
2048 1054
};
\addlegendentry{1000 eV, $N_i = 256$}

\addplot [only marks, dimgray85, mark=x, mark size=3, mark options={solid}]
table {%
32 15330
64 7848
128 4147
256 2202
512 1305
1024 873
2048 660
};

\addplot [only marks, dimgray85, mark=x, mark size=3, mark options={solid}]
table {%
32 30070
64 15035
128 7912
256 4131
512 2369
1024 1420
2048 1054
};

\addplot [only marks, dimgray85, mark=x, mark size=3, mark options={solid}]
table {%
32 18882
64 9563
128 5170
256 2736
512 1636
1024 1130
2048 926
};

\end{axis}

\end{tikzpicture}
\caption{Effect of computing resources: Total simulation time for 2000 timesteps as a function of the number of CPU cores for a system of $N_i \in$ \{108, 256\} hydrogen atoms at $\rho = 10$\,g/cm$^3$ and $T\in$\{1000, 5000\}\,eV.}
    \label{fig:scalability:c}
\end{figure}
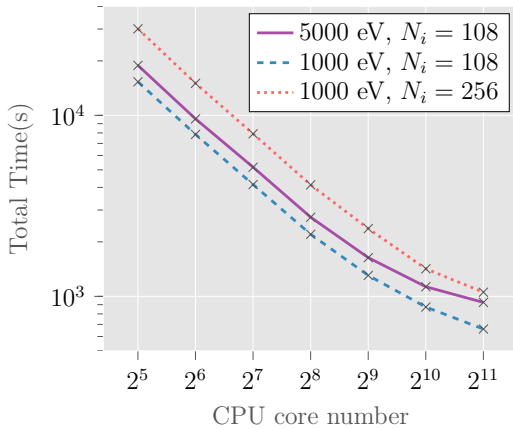

\begin{figure} [b!]
    \centering
\begin{tikzpicture}[xscale=0.8, yscale=0.8]

\definecolor{chocolate2267451}{RGB}{226,74,51}
\definecolor{dimgray85}{RGB}{85,85,85}
\definecolor{gainsboro229}{RGB}{229,229,229}
\definecolor{gray119}{RGB}{119,119,119}
\definecolor{lightpink255181184}{RGB}{255,181,184}
\definecolor{mediumpurple152142213}{RGB}{152,142,213}
\definecolor{sandybrown25119394}{RGB}{251,193,94}
\definecolor{steelblue52138189}{RGB}{52,138,189}
\definecolor{yellowgreen14218666}{RGB}{142,186,66}

\begin{axis}[
axis background/.style={fill=gainsboro229},
axis line style={white},
grid=major,
log basis x={2},
log basis y={10},
xmode=log,
ymode=log,
tick align=outside,
tick pos=left,
x grid style={white},
xlabel=\textcolor{dimgray85}{\(\displaystyle N_g\)},
xtick style={color=dimgray85},
y grid style={white},
ylabel=\textcolor{dimgray85}{Total Time(s)},
ymin=100, ymax=200000,
ytick style={color=dimgray85},
legend style={font=\small},
legend cell align=left,
legend pos=north west
]

\addplot [line width=1.32pt, violet!70]
table {%
262144 631
2097152 815
16777216 2170
134217728 14963
1073741824 116247
};
\addlegendentry{100 eV}

\addplot [line width=1.32pt, dashed, steelblue52138189]
table {%
262144 325
2097152 495
16777216 1483
134217728 7634
1073741824 65890
};
\addlegendentry{300 eV}

\addplot [line width=1.32pt, dotted, red!60]
table {%
262144 172
2097152 362
16777216 832
134217728 5349
1073741824 46928
};
\addlegendentry{900 eV}

\addplot [only marks, dimgray85, mark=x, mark size=3, mark options={solid}]
table {%
262144 631
2097152 815
16777216 2170
134217728 14963
1073741824 116247
};

\addplot [only marks, dimgray85, mark=x, mark size=3, mark options={solid}]
table {%
262144 325
2097152 495
16777216 1483
134217728 7634
1073741824 65890
};

\addplot [only marks, dimgray85, mark=x, mark size=3, mark options={solid}]
table {%
262144 172
2097152 362
16777216 832
134217728 5349
1073741824 46928
};

\end{axis}

\end{tikzpicture}
\caption{Effect of grid size: Total simulation time for 2000 timesteps as a function of the number $N_g$ of spatial/Fourier grid points for a system of 108 aluminum atoms at a compression of $4\rho_0$ and $T\in$\{100, 300, 900\} eV using 2048 CPU cores.}
    \label{fig:scalability:b}
\end{figure}
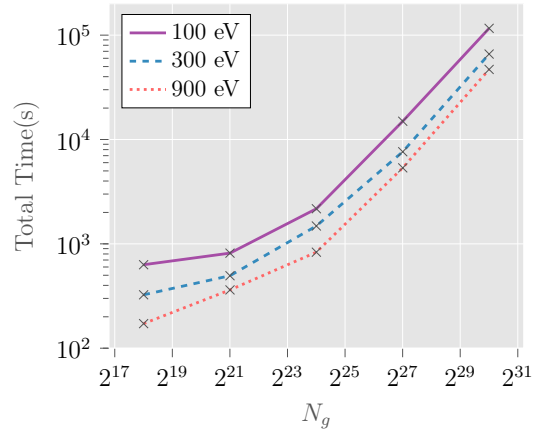

Fig. \ref{fig:scalability:c} displays the total simulation time as a function of the number of CPU cores under typical simulation conditions, spanning various temperatures and numbers of atoms. Scalability remains nearly ideal up to 512 cores, but decreases slightly beyond that point. This highlights the communication overhead affecting computation time, as parallelization efficiency drops when the number of subdomains increases.

Fig. \ref{fig:scalability:b} plots the total simulation time against the number of grid discretization points $N_g$ for different temperatures. These simulations were performed exclusively to measure performance, as the SCF convergence for the electronic density $n_e$ under these thermodynamic conditions had already been achieved with smaller grids ($64\times64\times64 = 2^{18}$). Since the complexity of the dominant algorithm in this simulation is governed by the Fourier transform $\mathcal{O}(N_g \log N_g)$, the computation time should theoretically increase slightly faster than linearly with the number of grid points, ignoring communication overhead. However, because the FFT implementation is highly optimized, the execution time appears to scale linearly with the number of grid points for the sizes tested, above $\approx 2^{24}$.

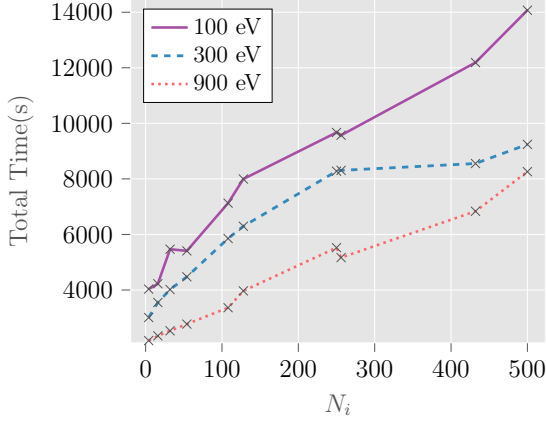
\begin{figure} [t!]
    \centering
    \begin{tikzpicture}[xscale=0.8, yscale=0.8]

\definecolor{chocolate2267451}{RGB}{226,74,51}
\definecolor{dimgray85}{RGB}{85,85,85}
\definecolor{gainsboro229}{RGB}{229,229,229}
\definecolor{gray119}{RGB}{119,119,119}
\definecolor{lightpink255181184}{RGB}{255,181,184}
\definecolor{mediumpurple152142213}{RGB}{152,142,213}
\definecolor{sandybrown25119394}{RGB}{251,193,94}
\definecolor{steelblue52138189}{RGB}{52,138,189}
\definecolor{yellowgreen14218666}{RGB}{142,186,66}

\begin{axis}[
axis background/.style={fill=gainsboro229},
axis line style={white},
grid=major,
tick align=outside,
tick pos=left,
x grid style={white},
xlabel=\textcolor{dimgray85}{\(\displaystyle N_i\)},
xmin=-19.5, xmax=523.5,
xtick style={color=dimgray85},
y grid style={white},
scaled y ticks=false,
ylabel=\textcolor{dimgray85}{Total Time(s)},
ymin=2100, ymax=14500,
ytick style={color=dimgray85},
legend style={font=\small},
legend cell align=left,
legend pos=north west
]

\addplot [line width=1.22pt, violet!70]
table {%
4 4033
16 4226
32 5469
54 5408
108 7127
128 7996
250 9671
256 9572
432 12188
500 14072
};
\addlegendentry{100 eV}

\addplot [line width=1.32pt, dashed, steelblue52138189]
table {%
4 3009
16 3550
32 4017
54 4476
108 5855
128 6295
250 8277
256 8307
432 8553
500 9243
};
\addlegendentry{300 eV}

\addplot [line width=1.22pt, dotted, red!60]
table {%
4 2180
16 2350
32 2532
54 2773
108 3363
128 3968
250 5524
256 5167
432 6835
500 8260
};
\addlegendentry{900 eV}

\addplot [only marks, dimgray85, mark=x, mark size=3, mark options={solid}]
table {%
4 4033
16 4226
32 5469
54 5408
108 7127
128 7996
250 9671
256 9572
432 12188
500 14072
};

\addplot [only marks, dimgray85, mark=x, mark size=3, mark options={solid}]
table {%
4 3009
16 3550
32 4017
54 4476
108 5855
128 6295
250 8277
256 8307
432 8553
500 9243
};

\addplot [only marks, dimgray85, mark=x, mark size=3, mark options={solid}]
table {%
4 2180
16 2350
32 2532
54 2773
108 3363
128 3968
250 5524
256 5167
432 6835
500 8260
};

\end{axis}

\end{tikzpicture}
\caption{Effect of system size: Total simulation time for 2000 timesteps as a function of the number $N_i$ of aluminum atoms at a compression of $4\rho_0$ and temperature $T\in$\{100, 300, 900\} eV using 256 CPU cores.}
    \label{fig:scalability:a}
\end{figure}

Fig. \ref{fig:scalability:a} presents the total simulation time as a function of the number of nuclei $N_i$ at different temperatures. The algorithmic complexity of the SCF loop is independent of $N_i$ (see Fig. \ref{fig:flowcomplexity}). However, $N_i$ plays a role in computing positions, velocities, $\vec{F}^{ie}$ forces, and in initializing the SCF loop. Furthermore, while the calculation of $\vec{F}^{ii}$ forces is proportional to $N_i^2$, this cost is negligible for the number of atoms used in these simulations. Consequently, the observed scalability with respect to the number of atoms $N_i$ is effectively linear.

 
\subsubsection{Trends with physical parameters}

Fig. \ref{fig:tempscaling} shows the total simulation time as a function of temperature for various compressions and numbers of atoms. An interesting trend emerges: the hotter the system, the shorter the time required to simulate it. This behavior is precisely the opposite of what is observed in KSDFT simulations, where higher temperatures lead to a greater number of occupied orbitals and, consequently, longer computation times.

In \textsc{Archê}, this reduction in execution time occurs because the self-consistent loop converges more rapidly. In fact, temperature has no direct impact on the number of operations required per iteration. However, at high temperatures, the electron density becomes more uniform than at low temperatures, allowing the algorithm to accurately converge in significantly fewer iterations.

\begin{figure} [t!]
    \centering
\begin{tikzpicture}[xscale=0.80, yscale=0.80]

\definecolor{chocolate2267451}{RGB}{226,74,51}
\definecolor{dimgray85}{RGB}{85,85,85}
\definecolor{gainsboro229}{RGB}{229,229,229}
\definecolor{gray119}{RGB}{119,119,119}
\definecolor{lightpink255181184}{RGB}{255,181,184}
\definecolor{mediumpurple152142213}{RGB}{152,142,213}
\definecolor{sandybrown25119394}{RGB}{251,193,94}
\definecolor{steelblue52138189}{RGB}{52,138,189}
\definecolor{yellowgreen14218666}{RGB}{142,186,66}

\begin{axis}[
axis background/.style={fill=gainsboro229},
axis line style={white},
grid=major,
tick align=outside,
tick pos=left,
x grid style={white},
xlabel=\textcolor{dimgray85}{\(\displaystyle \rho/\rho_0\)},
xmin=0.5, xmax=10.5,
xtick style={color=dimgray85},
y grid style={white},
ylabel=\textcolor{dimgray85}{Total Time(s)},
scaled y ticks=false,
ymin=2500, ymax=9000,
ytick style={color=dimgray85},
legend style={font=\small},
legend cell align=left,
legend pos=north west
]

\addplot [line width=1.32pt, violet!70] 
table {%
1 6009
2 6331
3 6117
4 6872
5 6779
6 6934
7 7099
8 8154
9 8082
10 7962
};
\addlegendentry{100 eV}

\addplot [line width=1.32pt, dashed, steelblue52138189] 
table {%
1 4697
2 4971
3 5239
4 5459
5 5899
6 5817
7 6159
8 6284
9 6407
10 6316
};
\addlegendentry{300 eV}

\addplot [line width=1.32pt, dotted, red!60] 
table {%
1 2695
2 3377
3 3213
4 3451
5 3374
6 3539
7 3963
8 3976
9 4023
10 3716
};
\addlegendentry{900 eV}

\addplot [only marks, dimgray85, mark=x, mark size=3, mark options={solid}]
table {%
1 6009
2 6331
3 6117
4 6872
5 6779
6 6934
7 7099
8 8154
9 8082
10 7962
};

\addplot [only marks, dimgray85, mark=x, mark size=3, mark options={solid}]
table {%
1 4697
2 4971
3 5239
4 5459
5 5899
6 5817
7 6159
8 6284
9 6407
10 6316
};

\addplot [only marks, dimgray85, mark=x, mark size=3, mark options={solid}]
table {%
1 2695
2 3377
3 3213
4 3451
5 3374
6 3539
7 3963
8 3976
9 4023
10 3716
};

\end{axis}

\end{tikzpicture}
\caption{Effect of density: Total simulation time for 2000 timesteps as a function of compression for 108 aluminum atoms at $T\in$\{100, 300, 900\} eV using 256 CPU cores.}
    \label{fig:densityscaling}
\end{figure}
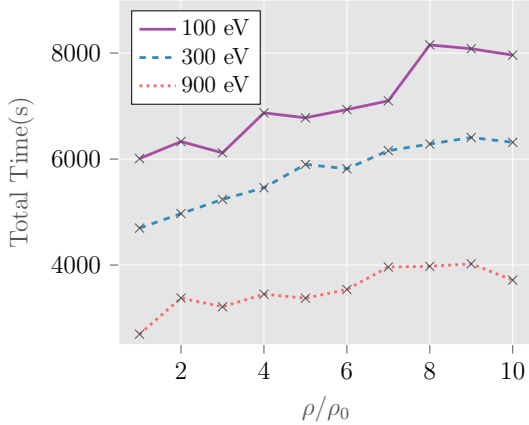

Fig. \ref{fig:densityscaling} illustrates the total simulation time as a function of density across different temperatures. As noted previously, the number of SCF iterations dominates the overall runtime. Density has no direct effect on the number of operations per iteration; nonetheless, as density increases, the SCF convergence requires slightly more iterations to be achieved.

\begin{figure} [t!]
    \centering
\begin{tikzpicture}[xscale=0.80, yscale=0.80]

\definecolor{chocolate2267451}{RGB}{226,74,51}
\definecolor{dimgray85}{RGB}{85,85,85}
\definecolor{gainsboro229}{RGB}{229,229,229}
\definecolor{gray119}{RGB}{119,119,119}
\definecolor{lightpink255181184}{RGB}{255,181,184}
\definecolor{mediumpurple152142213}{RGB}{152,142,213}
\definecolor{sandybrown25119394}{RGB}{251,193,94}
\definecolor{steelblue52138189}{RGB}{52,138,189}
\definecolor{yellowgreen14218666}{RGB}{142,186,66}

\begin{axis}[
axis background/.style={fill=gainsboro229},
axis line style={white},
grid=major,
tick align=outside,
tick pos=left,
x grid style={white},
xlabel=\textcolor{dimgray85}{\(\displaystyle Temperature\ (eV)\)},
xmin=-50., xmax=1050,
xtick style={color=dimgray85},
y grid style={white},
scaled y ticks=false,
ylabel=\textcolor{dimgray85}{Total Time(s)},
ymin=2500, ymax=11000,
ytick style={color=dimgray85}
]
\addplot [line width=1.32pt, violet!70]
table {%
10 10828
50 10850
100 10633
150 7966
200 7667
300 6932
400 6149
500 6237
750 5203
1000 4787
};
\addlegendentry{$4\rho_0, N_i = 256$}

\addplot [line width=1.32pt, dashed, steelblue52138189]
table {%
10 10248
50 10503
100 8332
150 7401
200 6931
300 5568
400 5028
500 4587
750 3858
1000 3332
};
\addlegendentry{$4\rho_0, N_i = 108$}

\addplot [line width=1.32pt, dotted, red!50]
table {%
10 6744
50 8712
100 7583
150 6634
200 5931
300 5139
400 4594
500 4378
750 3928
1000 3542
};
\addlegendentry{$8\rho_0, N_i = 108$}

\addplot [only marks, dimgray85, mark=x, mark size=3, mark options={solid}]
table {%
10 10828
50 10850
100 10633
150 7966
200 7667
300 6932
400 6149
500 6237
750 5203
1000 4787
};

\addplot [only marks, dimgray85, mark=x, mark size=3, mark options={solid}]
table {%
10 10248
50 10503
100 8332
150 7401
200 6931
300 5568
400 5028
500 4587
750 3858
1000 3332
};

\addplot [only marks, dimgray85, mark=x, mark size=3, mark options={solid}]
table {%
10 6744
50 8712
100 7583
150 6634
200 5931
300 5139
400 4594
500 4378
750 3928
1000 3542
};
\end{axis}

\end{tikzpicture}
\caption{Effect of temperature: Total simulation time for 2000 timesteps as a function of temperature for $N_i \in$\{108, 256\} aluminum atoms at compressions of \{4, 8\}$\rho_0$ using 256 CPU cores.}
    \label{fig:tempscaling}
\end{figure}
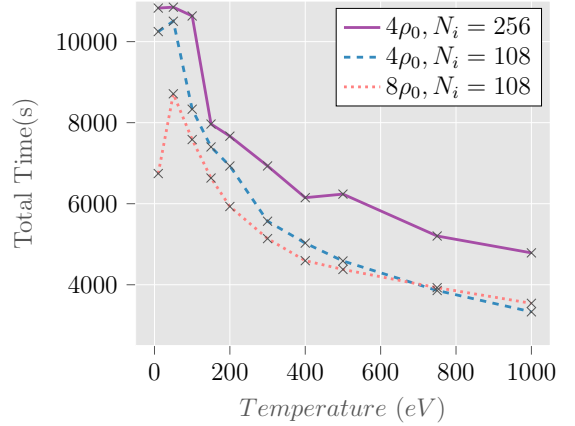

\subsection{GPU speedup}

To fully exploit the A100 processors, the most computationally demanding parts of \textsc{Archê} were ported to GPUs. In the CPU version, the majority of the runtime is spent on the Fourier transform due to its individual cost and the high number of calls. Recognizing that the CUDA toolkit provides a highly efficient FFT library called\linebreak CuFFT, we replaced our CPU library with\linebreak NVIDIA's single-GPU implementation, which generated most of the results presented here. Since then, we have integrated the HeFFTe library \cite{HEFFTEfirst} using the CuFFT backend. This enables multi-GPU execution and is mandatory when handling grid sizes of one billion points or more.

To maximize GPU acceleration, it is critical to minimize memory transfers between the host (CPU) and the device (GPU). For this reason, when computing the electronic density, all fields associated with the grid are allocated and evaluated directly on the GPU. By keeping all grid arrays on the GPU, memory transfers per timestep are reduced to just two operations: copying the nuclear positions (input) and retrieving the\linebreak electron-nucleus forces $\vec{F}^{ie}$ (output), which amounts to $2 \times 3 \times N_i$ floating-point numbers. Consequently, the entire SCF loop --- including its initialization and the final force calculations, which account for up to 95\% of the CPU computation time --- is executed on the GPU with minimal data movement.

\begin{figure} [t!] 
    \centering
\begin{tikzpicture}[xscale=0.8, yscale=0.8]

\definecolor{chocolate2267451}{RGB}{226,74,51}
\definecolor{dimgray85}{RGB}{85,85,85}
\definecolor{gainsboro229}{RGB}{229,229,229}
\definecolor{gray119}{RGB}{119,119,119}
\definecolor{lightpink255181184}{RGB}{255,181,184}
\definecolor{mediumpurple152142213}{RGB}{152,142,213}
\definecolor{sandybrown25119394}{RGB}{251,193,94}
\definecolor{steelblue52138189}{RGB}{52,138,189}
\definecolor{yellowgreen14218666}{RGB}{142,186,66}

\begin{axis}[
axis background/.style={fill=gainsboro229},
axis line style={white},
grid=major,
tick align=outside,
tick pos=left,
x grid style={white},
xlabel=\textcolor{dimgray85}{\(\displaystyle Temperature\ (eV)\)},
xtick style={color=dimgray85},
y grid style={white},
ylabel=\textcolor{dimgray85}{Total time ratio CPU/GPU},
ytick style={color=dimgray85},
legend style={font=\small},
legend cell align=left,
legend pos=north east
]
\addplot [line width=1.32pt, violet!70]
table {%
10 11.12
50 11.39
100 12.53
150 10.04
200 10.21
300 9.72
400 8.93
500 9.31
750 8.24
1000 7.89
};
\addlegendentry{$4\rho_0, N_i = 256$}

\addplot [line width=1.32pt, dashed, steelblue52138189]
table {%
10 15.15
50 14.03
100 13.53
150 13.54
200 13.82
300 12.20
400 11.74
500 11.29
750 10.50
1000 9.81
};
\addlegendentry{$4\rho_0, N_i = 108$}

\addplot [line width=1.32pt, dotted, red!60]
table {%
10 13.2
50 14.06
100 12.84
150 12.19
200 11.97
300 11.38
400 10.87
500 10.73
750 10.44
1000 9.80
};
\addlegendentry{$8\rho_0, N_i = 108$}

\addplot [only marks, dimgray85, mark=x, mark size=3, mark options={solid}]
table {%
10 11.12
50 11.39
100 12.53
150 10.04
200 10.21
300 9.72
400 8.93
500 9.31
750 8.24
1000 7.89
};

\addplot [only marks, dimgray85, mark=x, mark size=3, mark options={solid}]
table {%
10 15.15
50 14.03
100 13.53
150 13.54
200 13.82
300 12.20
400 11.74
500 11.29
750 10.50
1000 9.81
};

\addplot [only marks, dimgray85, mark=x, mark size=3, mark options={solid}]
table {%
10 13.2
50 14.06
100 12.84
150 12.19
200 11.97
300 11.38
400 10.87
500 10.73
750 10.44
1000 9.80
};

\end{axis}

\end{tikzpicture}
\caption{GPU speedup relative to temperature: Ratio of the total simulation time (2000 timesteps) on 256 CPU cores to that on 1 GPU, as a function of temperature, for $N_i \in \{108, 256\}$ aluminum atoms at compressions of \{4, 8\} $\rho_0$.}
    \label{fig:gpu:temperature}
\end{figure}
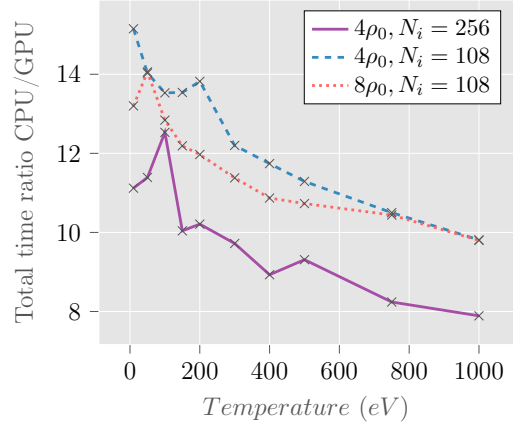

In the results presented below, while GPU efficiency varies depending on the specific conditions, running \textsc{Archê} on a GPU consistently provides a massive advantage over a CPU. Currently, the primary limitation is the memory capacity of a single GPU. To simulate larger systems requiring larger grids, we employ the multi-GPU version utilizing domain decomposition, where each subdomain is processed by a single GPU.

Figure~\ref{fig:gpu:temperature} presents the ratio of CPU to GPU simulation times as a function of temperature for various compressions and system sizes. GPU efficiency slightly decreases at higher temperatures because fewer iterations are required for the electronic density to converge. This relatively increases the weight of the code sections still executed on the CPU, such as the Ewald summation and the updates of particle positions and velocities. Nevertheless, even in the least favorable scenario of this benchmark, a single GPU outperforms 256 CPU cores by a factor of roughly eight.

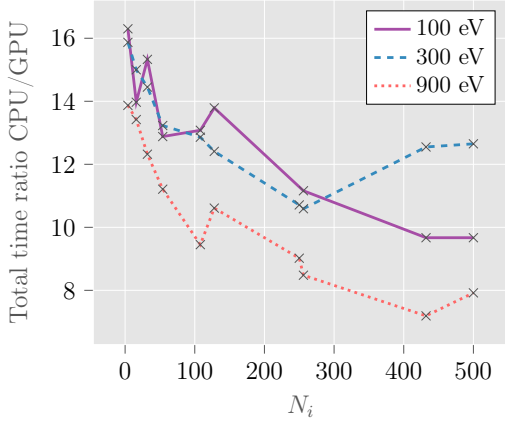
\begin{figure}[t!]
    \centering
\begin{tikzpicture}[xscale=0.8, yscale=0.8]

\definecolor{chocolate2267451}{RGB}{226,74,51}
\definecolor{dimgray85}{RGB}{85,85,85}
\definecolor{gainsboro229}{RGB}{229,229,229}
\definecolor{gray119}{RGB}{119,119,119}
\definecolor{lightpink255181184}{RGB}{255,181,184}
\definecolor{mediumpurple152142213}{RGB}{152,142,213}
\definecolor{sandybrown25119394}{RGB}{251,193,94}
\definecolor{steelblue52138189}{RGB}{52,138,189}
\definecolor{yellowgreen14218666}{RGB}{142,186,66}

\begin{axis}[
axis background/.style={fill=gainsboro229},
axis line style={white},
grid=major,
tick align=outside,
tick pos=left,
x grid style={white},
xlabel=\textcolor{dimgray85}{\(\displaystyle N_i\)},
xtick style={color=dimgray85},
y grid style={white},
ylabel=\textcolor{dimgray85}{Total time ratio CPU/GPU},
ytick style={color=dimgray85},
legend style={font=\small},
legend cell align=left,
legend pos=north east
]

\addplot [line width=1.32pt, violet!70]
table {%
4 16.3
16 13.97
32 15.33
54 12.88
108 13.08
128 13.79
256 11.16
432 9.67
500 9.67
};
\addlegendentry{100 eV}

\addplot [line width=1.32pt, dashed, steelblue52138189]
table {%
4 15.87
16 15
32 14.45
54 13.23
108 12.86
128 12.41
250 10.71
256 10.59
432 12.55
500 12.65
};
\addlegendentry{300 eV}

\addplot [line width=1.32pt, dotted, red!60]
table {%
4 13.87
16 13.42
32 12.32
54 11.21
108 9.45
128 10.61
250 9.02
256 8.48
432 7.19
500 7.92
};
\addlegendentry{900 eV}

\addplot [only marks, dimgray85, mark=x, mark size=3, mark options={solid}]
table {%
4 16.3
16 13.97
32 15.33
54 12.88
108 13.08
128 13.79
256 11.16
432 9.67
500 9.67
};

\addplot [only marks, dimgray85, mark=x, mark size=3, mark options={solid}]
table {%
4 15.87
16 15
32 14.45
54 13.23
108 12.86
128 12.41
250 10.71
256 10.59
432 12.55
500 12.65
};

\addplot [only marks, dimgray85, mark=x, mark size=3, mark options={solid}]
table {%
4 13.87
16 13.42
32 12.32
54 11.21
108 9.45
128 10.61
250 9.02
256 8.48
432 7.19
500 7.92
};

\end{axis}

\end{tikzpicture}
\caption{GPU speedup relative to system size: Ratio of the total simulation time (2000 timesteps) on 256 CPU cores to that on 1 GPU, as a function of the number of aluminum atoms $N_i$ at a compression of $4\rho_0$ and temperatures of \{100, 300, 900\} eV.}
    \label{fig:gpu:nucleus}
\end{figure}

Fig. \ref{fig:gpu:nucleus} shows the ratio of CPU to GPU simulation times as a function of the number of atoms in the simulation box for different temperatures. The general trend suggests a slight decrease in efficiency as the number of atoms $N_i$ grows. This can be attributed to increased CPU-GPU data transfers and a larger fraction of time spent on CPU-side computations. However, the curve corresponding to simulations at 300 eV does not exhibit a clear, monotonic trend.

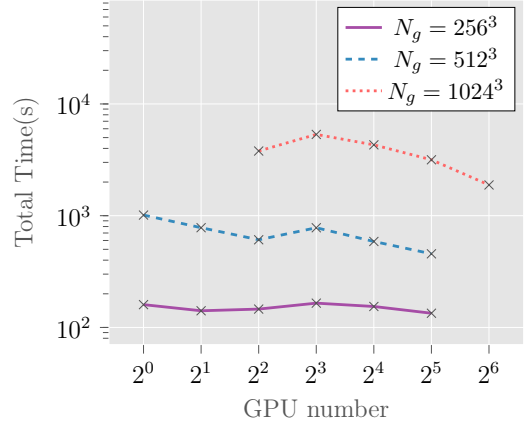
\begin{figure}
\centering
\begin{tikzpicture}[xscale=0.8, yscale=0.8]

\definecolor{chocolate2267451}{RGB}{226,74,51}
\definecolor{dimgray85}{RGB}{85,85,85}
\definecolor{gainsboro229}{RGB}{229,229,229}
\definecolor{gray119}{RGB}{119,119,119}
\definecolor{lightpink255181184}{RGB}{255,181,184}
\definecolor{mediumpurple152142213}{RGB}{152,142,213}
\definecolor{sandybrown25119394}{RGB}{251,193,94}
\definecolor{steelblue52138189}{RGB}{52,138,189}
\definecolor{yellowgreen14218666}{RGB}{142,186,66}

\begin{axis}[
axis background/.style={fill=gainsboro229},
axis line style={white},
grid=major,
ymax=85000,
log basis x={2},
log basis y={10},
xmode=log,
ymode=log,
tick align=outside,
tick pos=left,
x grid style={white},
xlabel=\textcolor{dimgray85}{GPU number},
xtick style={color=dimgray85},
y grid style={white},
ylabel=\textcolor{dimgray85}{Total Time(s)},
ytick style={color=dimgray85},
legend style={font=\small},
]

\addplot [line width=1.32pt, violet!70]
table {%
1 160
2 141
4 146
8 165
16 154
32 134
};
\addlegendentry{$N_g = 256^3$}

\addplot [line width=1.32pt, dashed, steelblue52138189]
table {%
1 1013
2 781
4 609
8 780
16 588
32 456
};
\addlegendentry{$N_g = 512^3$}

\addplot [line width=1.32pt, dotted, red!60]
table {%
4 3792
8 5337
16 4306
32 3165
64 1884
};
\addlegendentry{$N_g = 1024^3$}

\addplot [only marks, dimgray85, mark=x, mark size=3, mark options={solid}]
table {%
1 160
2 141
4 146
8 165
16 154
32 134
};

\addplot [only marks, dimgray85, mark=x, mark size=3, mark options={solid}]
table {%
1 1013
2 781
4 609
8 780
16 588
32 456
};

\addplot [only marks, dimgray85, mark=x, mark size=3, mark options={solid}]
table {%
4 3792
8 5337
16 4306
32 3165
64 1884
};

\end{axis}

\end{tikzpicture}
\caption{Total simulation time for 500 timesteps as a function of the number of GPUs for 216 oxygen atoms at 20 g/cm$^3$ and 100 eV for grid sizes $N_g \in \{256^3, 512^3, 1024^3\}$.}
\label{fig:multigpu:strongscaling}
\end{figure}

Fig. \ref{fig:multigpu:strongscaling} illustrates the strong scaling of the multi-GPU version of the code across different grid sizes. This version introduces inter-GPU communications, which inevitably degrade per\-for\-man\-ce. Because the simulations were run on nodes containing four GPUs each, communications between different nodes penalize performance even further. For this reason, it is always preferable to use a single GPU when its memory is sufficient to hold the simulation grid. The results indicate that scaling for the $256^3$ grid is disappointing; ideally, a satisfactory scaling curve would resemble Fig. \ref{fig:scalability:c}. The explanation is straightforward: GPUs excel at processing massive blocks of data, whereas inter-GPU communications --- especially inter-node ones --- carry a very high latency cost on our architecture.

\section{Conclusion}
\label{conclusion}

We have presented \textsc{Archê}, an OFMD code designed to compute equations of state (EOS) for plasmas, offering an optimal balance between accuracy and efficiency.

The core design is detailed alongside the governing equations, which are presented as closely as possible to their implementation in the source code. 
Unlike other OFMD codes, \textsc{Archê} relies on a self-consistent field (SCF) approach to calculate the electron density, rather than minimizing the free energy via a conjugate gradient method. This choice enabled the implementation of two algorithms that accelerate SCF convergence by up to a factor of six. Complete derivations for both improvements are provided. 

The first improvement to the SCF scheme involves using the converged density field from the previous MD timestep to extract an average one-center density profile. This profile is then superposed onto the new nuclear positions to initialize the density for the current SCF cycle.

The second improvement adapts an algorithm originally proposed for average-atom models by More \textit{et al.} \cite{More1988} for use in molecular dynamics. At each SCF iteration, the initial and active densities are mixed to accelerate convergence. The mixing proportion is dynamically determined to minimize a second-order approximation of the free energy.

We also detailed the introduction of a norm-conserving pseudopotential, derived from an average atom model, to describe the electron-nucleus interaction. Particular emphasis was placed on the energy correction required to yield an accurate EOS.

Using the simple Thomas-Fermi kinetic functional, the code was validated for aluminum by comparing its results with Kohn-Sham molecular dynamics simulations performed using \textsc{Abinit Extended} \cite{BlanchetAluminum2022}. At high temperatures (above 10 eV), the pressure and internal energy values show agreement typically within 1\% to 5\%. Larger discrepancies appear at lower temperatures; these can be mitigated by incorporating exchange and correlation corrections, as well as by utilizing semi-local or non-local kinetic energy functionals. These refinements will be explored in future work.

\textsc{Archê}'s performance was evaluated across various architectures, including multi-CPU, single-GPU, and multi-GPU setups. For EOS production, the single-GPU configuration proves to be the most efficient, running an order of magnitude faster than a 256-core CPU architecture. While the multi-GPU architecture allows for larger memory allocation, it currently does not offer the same scalability as its multi-CPU counterpart.

Performance trends as a function of both numerical and physical parameters were analyzed and explained in light of the code's algorithmic complexity. In summary, \textsc{Archê} exhibits overall linear complexity with respect to the number of atoms and the number of real and reciprocal grid points. Execution time is weakly dependent on density; however, interestingly, it decreases as temperature increases --- a stark contrast to the behavior of Kohn-Sham orbital-based simulations.

We believe that Archê establishes a flexible software foundation for testing state-of-the-art kinetic energy functionals at finite temperatures. This addresses the need for accurate EOS and transport coefficients.

\section*{Declaration of competing interest}
The authors declare that they have no competing financial interests or personal relationships that could have appeared to influence the work reported in this paper. AI and large language models (LLMs) were not systematically used in the writing of this paper, and were strictly excluded from the research process.

\section*{Data availability}

Data other than cited are confidential.

\section*{Acknowledgments}
We would like to acknowledge Gilles Kluth, Gwenaël Salin, Etienne Jaupard, Mikael Tacu, Vanina Recoules, and Gérald Faussurier for their valuable advice and fruitful discussions.

\bibliographystyle{elsarticle-num}
\bibliography{common.bib}

\end{document}